\documentclass{article}
\usepackage[T1]{fontenc}
\usepackage{iclr2027_conference,times}
\iclrfinalcopy
\usepackage{amsmath,amssymb,graphicx,booktabs}
\usepackage{hyperref,url}
\hypersetup{hidelinks,pdfauthor={Kehan Lu, Tianhao Chu, AIERSI TUERHONG, Yiwen Sun, Huan Luo, Fang Fang, Si Wu},pdftitle={Oscillatory Phase Coding Emerges in Adaptive RNNs for Flexible Working Memory}}
\newcommand{\R}{\mathbb{R}}

\newcommand{\figref}[1]{Figure~\ref{#1}}
\newcommand{\secref}[1]{Section~\ref{#1}}
\newcommand{\appref}[1]{Appendix~\ref{#1}}
\newcommand{\arnn}{A-RNN}

\title{Oscillatory Phase Coding Emerges in\\Adaptive RNNs for Flexible Working Memory}
\author{%
\parbox{\dimexpr\textwidth-2\tabcolsep\relax}{%
\raggedright
\normalsize\bfseries
Kehan Lu\textsuperscript{1,\ensuremath{\dagger}}\quad
Tianhao Chu\textsuperscript{2,\ensuremath{\dagger},*}\quad
AIERSI TUERHONG\textsuperscript{3}\quad
Yiwen Sun\textsuperscript{3}\\[4pt]
Huan Luo\textsuperscript{2}\quad
Fang Fang\textsuperscript{2}\quad
Si Wu\textsuperscript{2,*}%
}}
\newcommand{\arxivauthornotes}{%
\textsuperscript{1}\,Academy for Advanced Interdisciplinary Studies, Peking University, Beijing, China.\\
\textsuperscript{2}\,School of Psychological and Cognitive Sciences, Peking University, Beijing, China.\\
\textsuperscript{3}\,Peking University, Beijing, China.\\
\textsuperscript{\ensuremath{\dagger}}\,Co-first authors.\quad
\textsuperscript{*}\,Corresponding authors.\\
Email: \href{mailto:siwu@pku.edu.cn}{\texttt{siwu@pku.edu.cn}};\quad \href{mailto:chutianhao@pku.edu.cn}{\texttt{chutianhao@pku.edu.cn}}%
}
\begin{document}
% Keep figure-to-text gaps compact rather than stretching them to fill the page.
\raggedbottom
\setlength{\textfloatsep}{12pt plus 2pt minus 2pt}
\maketitle
\fancyhead{}
\renewcommand{\headrulewidth}{0pt}
\begingroup
\renewcommand{\thefootnote}{}
\makeatletter
\long\def\@makefntext#1{\noindent#1}
\makeatother
\footnotetext{\arxivauthornotes}
\endgroup
\begin{abstract}
Oscillatory phase coding (OPC) is widely observed in neural systems and has been implicated in cognitive functions such as memory and sequence processing. However, its computational role and neural substrates remain largely unclear. Here, we studied OPC by training recurrent neural networks (RNNs) with adaptation to perform sequence working memory tasks with variable delays. After training, we found that OPC emerges in the RNN spontaneously: the memorized items recur at distinct phases of a shared rhythm, with oscillation amplitude encoding item identity and phase encoding serial order. Ablation studies identified three key elements required for OPC: adaptation in neural dynamics, variable delay in memory retention, and flexibility in reading out information. Neuron-type analysis revealed the emergence of content and time cells in the trained RNN, and they process item identity and sequence order separately. Recurrent connectivity analysis revealed phase-organized connections among time cells and structured connections within digit-selective content-cell groups. Inspired by these findings, we proposed a neural circuit model which reproduces OPC through interactions between content units and an adaptive time ring. Together, this study reveals how OPC arises from network structures and task demands, offers insights into the implementation and computational mechanisms of OPC in biological systems, and inspires us to develop computational models for human-like cognitive functions.
\end{abstract}

\section{Introduction}
Oscillation is a pervasive feature of neural activity across brain regions and is thought to play important functional roles \citep{buzsaki2004,lisman2013,malhotra2012,skaggs1996}. A promising theory called oscillatory phase coding (OPC) suggests that neural oscillations provide a reference framework for leveraging oscillation phases to encode information \citep{skaggs1996,buzsaki2004,lisman2013}. Consistent with this theory, experimental studies have shown that hippocampal place cells exhibit theta phase precession, with different place cells activated at different theta phases \citep{okeefe1993,skaggs1996}. Similarly, cortical circuits exhibit cross-frequency coupling, in which high-frequency gamma activity is modulated by the phase of a slower theta oscillation \citep{canolty2006}.

\begin{figure}[t]
\centering
\includegraphics[width=\linewidth]{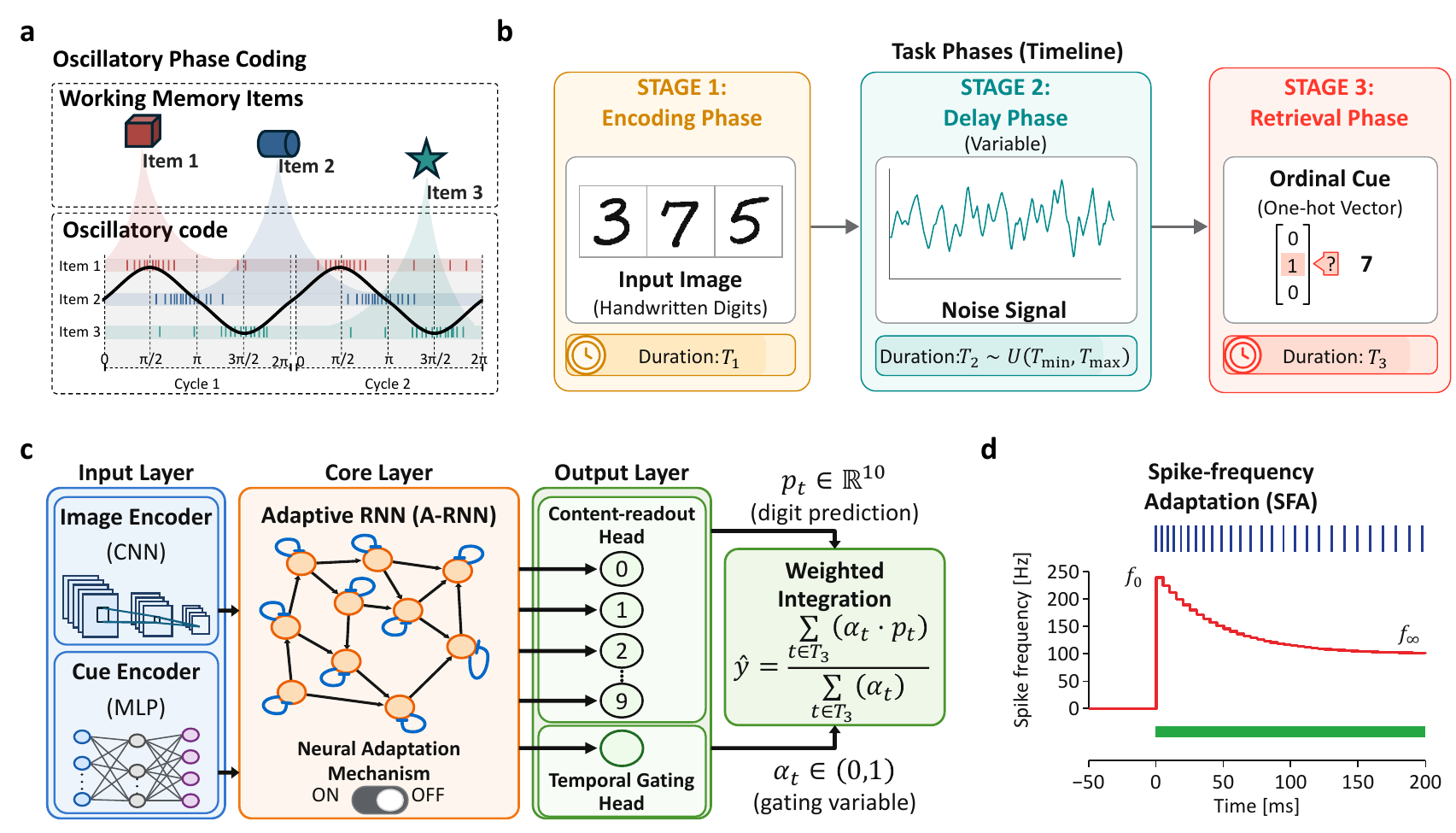}
\caption{\textbf{Task and model of sequence working memory.} (a) Illustration of OPC, where memory items recur at distinct phases across successive oscillation cycles. (b) Memory items (digits) are loaded in sequence, maintained across a variable delay, and retrieved by an ordinal cue. Three items are illustrated. (c) An adaptive RNN receives image and cue inputs. The content head outputs digit scores $\mathbf p_t$, while the temporal gate $\alpha_t$ controls the information readout. (d) Illustration of spike-frequency adaptation, by which neuronal activity declines in response to a sustained input.}
\label{fig:task}
\end{figure}

OPC has been proposed as a neural code for sequence working memory, in which the neural system needs to memorize both the identity and serial order of each item. In the theta--gamma coding framework developed by Lisman and colleagues, neuronal ensembles representing individual items are reactivated in successive gamma cycles nested within a slower theta rhythm \citep{lisman1995,lisman2013}. Repeated reactivation sustains item representations, while their relative phases within each theta cycle encode serial orders (\figref{fig:task}a). It remains, however, unclear how neural circuits generate and sustain OPC.

Training recurrent neural networks (RNNs) on cognitive tasks has become an increasingly popular approach to study brain functions \citep{yangwang2020}. Varying task demands, network architectures, and biological constraints in training RNNs reveals the conditions under which neural representations emerge and offers insight into the computational mechanisms of neural systems \citep{banino2018,Yang2019,chu2026unfolding}. Here, we applied this approach to investigate the emergence and computational role of OPC in sequence working memory.

Specifically, we trained recurrent neural networks with adaptation (\arnn{}s) to carry out a sequence working memory task with flexible delays. Adaptation was motivated by previous studies showing that spike-frequency adaptation is necessary to generate oscillatory dynamics associated with the hippocampal theta phase precession \citep{chu2024,ji2025}. Performing this task requires the network to memorize the sequence of items (handwritten digits used in the present study), maintain them across a variable delay interval, before recalling one of them specified by the ordinal cue. Here, the variable delay,
which prevents reliance on a fixed retrieval time,
imposes a strong constraint on the learned solution of the network, reflecting the flexibility of human memory.

After training, we found that the \arnn{} spontaneously develops oscillatory phase coding, in that the memorized items are reactivated at distinct phases in an internally generated rhythm, with oscillation amplitude encoding item identity and phase encoding serial order. This coding scheme provides disentangled representations for item identity and serial order. When the ordinal cue is presented, it modulates temporal gating to selectively read out the cued item.

We carried out ablation studies and identified neural adaptation, variable delay, and flexible readout timing as necessary for generating stable OPC. In particular, adaptation contributes to generating oscillation, and its learnable strength allows the network to adjust the oscillation frequency. We also showed that adaptive RNNs learn faster and attain higher final accuracy than non-adaptive ones (\appref{app:s1}). Variable delays contribute to inducing phase coding, as this is a stable solution generalizable across varied time delays. The flexible readout timing provides a suitable strategy to extract information from phase coding, as in OPC the item information does not appear at a fixed time. Together, these results suggest that OPC emerges as an optimized solution under the demand of flexible working memory and the support of adaptive neural dynamics.

We further carried out analyses to categorize neuron types in the trained \arnn{}s, and we identified two main neuron types: content and time cells. Content cells encode the item identity and their phases track the serial position of the item, while time cells form a traveling wave which provides a shared temporal reference for all memorized items. Time cells exhibit a recurrent connectivity kernel with an approximately cosine shape on the phase ring, while content cells show more organized recurrent connectivity within digit-selective groups than across groups.

Inspired by these findings, we proposed a neural circuit model that reproduces OPC through interactions between content units and an adaptive time ring, providing a simple account of how OPC could operate in biological circuits.

Our study also generated experimentally testable predictions regarding the dissociation between memory content and serial order in biological circuits. According to our model, time cells maintain consistent oscillatory phases across sequences, whereas content cells remain phase-locked to the reactivation of their associated items. Thus, our model predicts that changing an item's serial position will shift content-cell phases while preserving the temporal reference provided by time cells.

\section{Task Design and Model Architecture}
\label{sec:method}

\subsection{Task Design}
The working memory task we considered follows the conventional protocol of human psychophysical experiments, which consists of three stages, as illustrated in \figref{fig:task}b.

\noindent\textbf{1) Encoding stage (duration: $T_1$).} A sequence of $N$ memory items (handwritten digits) is presented at fixed intervals, with only one item visible at a time. The network must memorize the identity of items and their serial order.

\noindent\textbf{2) Delay stage (duration: $T_2$).} Task-relevant inputs are removed, and the network receives only background noise. The network must memorize the item sequence during this period, and the duration $T_2$ is variable, which is sampled uniformly from $\mathcal U(T_{\min},T_{\max})$ across trials.

\noindent\textbf{3) Retrieval stage (duration: $T_3$).} An ordinal cue specifying the serial order of the item to be recalled is presented, and the network needs to report the identity of the corresponding item.

\subsection{Model Architecture}
The model comprises three parts, as illustrated in \figref{fig:task}c:

\noindent\textbf{1) Sensory Encoders:}
\begin{itemize}
\item \textbf{Visual Encoder (CNN):} A convolutional neural network extracts visual features from handwritten digits during the encoding stage and projects them to the RNN.
\item \textbf{Cue Encoder (MLP):} A multilayer perceptron encodes the ordinal cue during the retrieval stage and projects it to the RNN.
\end{itemize}

\noindent\textbf{2) Adaptive Recurrent Neural Network (A-RNN):} The recurrent network integrates external inputs and the activities of hidden neurons, with adaptation in the dynamics of single neurons (\figref{fig:task}d).

Let $\mathbf x_t^{(s)}\in\R^{D_s}$ and $\mathbf x_t^{(c)}\in\R^{D_c}$ denote the visual and cue inputs at time $t$, respectively. The activity of $H$ hidden neurons is denoted as $\mathbf h_t\in\R^H$, and their adaptation states as $\boldsymbol\theta_t\in\R^H$. The network dynamics are given by
\begin{align}
\mathbf z_t &= \operatorname{CNN}\!\left(\mathbf x_t^{(s)}\right)
 +\operatorname{MLP}\!\left(\mathbf x_t^{(c)}\right)
 +W\mathbf h_{t-1}+\mathbf b,\label{eq:drive}\\
\mathbf h_t &= (1-\alpha_h)\mathbf h_{t-1}
 +\alpha_h\tanh\!\bigl(\operatorname{LN}(\mathbf z_t)\bigr)
 -\boldsymbol\theta_{t-1},\label{eq:activity}\\
\boldsymbol\theta_t &= (1-\alpha_\theta)\boldsymbol\theta_{t-1}
 +\alpha_\theta(\mathbf m\odot\mathbf h_t),\label{eq:adaptation}
\end{align}
where $W$ denotes the recurrent connectivity, $\mathbf b$ a bias vector, and $\odot$ the element-wise multiplication. The parameters $\alpha_h,\alpha_\theta\in(0,1]$ govern the update rates of neural activity and adaptation, respectively, and $\operatorname{LN}$ denotes layer normalization across hidden neurons.

Spike-frequency adaptation is widely observed in biological neurons, whose firing rates decline under sustained stimulation \citep{madison1984,schwindt1989}. One mechanism involves slowly accumulating potassium currents which provide negative feedback to reduce neuronal excitability \citep{madison1984,schwindt1989}. In our model, these adaptive dynamics are captured by the adaptation state $\boldsymbol\theta_t$, which evolves more slowly than the neural activity ($\alpha_\theta<\alpha_h$) and feeds back to the network dynamics through the term $-\boldsymbol\theta_{t-1}$. The nonnegative parameter $m_i$ controls the adaptation strength and is learned jointly with connection weights.

\noindent\textbf{3) Dual-readout Pathways:} Two linear readouts of neural activity jointly produce the network output: one reads out the scores of item identity over time, and the other controls the timing of reading out information specified by the ordinal cue. This design draws on the studies of output gating in working memory \citep{chatham2014} and temporal attention \citep{denison2021}.
\begin{itemize}
\item \textbf{Content Head:} This pathway reads out the item identity from neural activity, producing a vector of item scores (logits) $\mathbf p_t$. $\mathbf p_t$ varies rhythmically over time as a consequence of OPC.
\item \textbf{Gating Head:} This pathway reads out
the timing of extracting the target item specified by the ordinal cue. The timing is represented by
a gating variable $\alpha_t\in[0,1]$, which peaks at the phase when the oscillation of the target item reaches its maximum.
\end{itemize}
The network output is given by:
\begin{equation}
\widehat{\mathbf y}=\frac{\sum_{t\in T_3}\alpha_t\mathbf p_t}
{\sum_{t\in T_3}\alpha_t}.
\label{eq:readout}
\end{equation}
We see that in the framework of OPC, the multiplication
$\alpha_t\mathbf p_t$ has the largest element for the cued item (see illustration in \figref{fig:code}e). Hence, the largest element of $\widehat{\mathbf y}$ predicts the item identity.

In our model, all network parameters are trained jointly by minimizing the cross-entropy loss between $\widehat{\mathbf y}$ and the identity of the cued item $\mathbf y_{\mathrm{true}}$. Input pathways and readout calculations are detailed in \appref{app:readout}; training settings are provided in \appref{app:optimization}.

\section{Results}
\label{sec:results}
\subsection{Oscillatory phase coding emerges in trained RNNs}
\label{sec:phasecode}
\begin{figure}[t]
\centering
\includegraphics[width=0.9\linewidth]{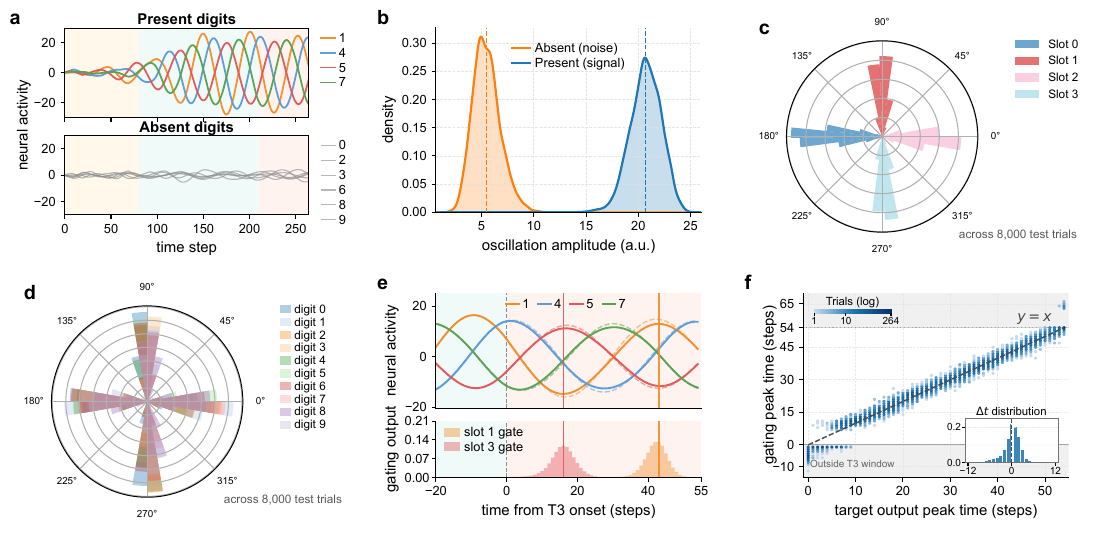}
\caption{\textbf{Oscillation phase coding emerges in the trained A-RNN.} (a) Rhythmic oscillations for present items (digits) exhibiting distinct phases (colored). No oscillation for absent items (gray). Shading indicates the task stages. (b) Oscillation-amplitude distributions for present and absent items. (c,d) Phase distributions over trials are separated by serial orders (c) but overlapped by item identities (d). (e) Changing the ordinal cue shifts the temporal gate toward the target item's phase while preserving the content encoding. (f) Temporal gate and target item phase align with each other across correctly retrieved trials. The inset shows circular timing differences. Analysis details appear in \appref{app:phase}.}
\label{fig:code}
\end{figure}

After training, the \arnn{} performs the sequence working memory task with high accuracy, and it develops OPC spontaneously. Indeed, projecting neural activity onto the content-readout head predicts the identity of the cued item. During the delay period, the stored items oscillate at the same rhythm but occupy distinct phases (\figref{fig:code}a). This temporal organization allows the sequence to be represented through recurring item-associated activations within each cycle.

We found that oscillation amplitude carries the ``what'' information. Present-digit channels exhibit large-amplitude oscillations, while absent-digit channels remain largely silent, producing separated amplitude distributions (\figref{fig:code}b). This amplitude organization correlates with retrieval performance. On average, incorrect trials have lower present-digit amplitudes and higher absent-digit amplitudes than correct trials, reducing the separation between present and absent digits (\appref{app:amplitude-performance}). 

We further confirmed that oscillation phase carries the ``where'' information, i.e., the item's order in the sequence. Grouping oscillation phases over trials according to their serial positions displays concentrated, well separated phase sectors (\figref{fig:code}c), while grouping phases according to item identities produces overlapped distributions (\figref{fig:code}d). These observations suggest
that the RNN learns a phase-coding scaffold to represent 
the sequence order, binding each item to its position across trials. 

In the retrieval stage, the network reads out the identity of the target item by combining the outputs of the content and gating heads. 
Specifically, the temporal gate learns to select the oscillation phase of the cued item. This is evident in that
for the same sequence, changing the ordinal cue largely preserves the content-related oscillation activities but shifts the temporal gate toward the phase of the newly cued item (\figref{fig:code}e). We showed that across correctly retrieved trials with variable delays, the temporal gate is always closely aligned with the phase of the target item (\figref{fig:code}f). This conclusion is validated by 
independent evaluations of trained RNNs with varying seeds and configurations (\appref{app:robust-opc}; \figref{fig:multiseed-robustness}).

\subsection{Conditions for the emergence of OPC}
\label{sec:conditions}
\begin{figure}[t]
\centering
\includegraphics[width=\linewidth]{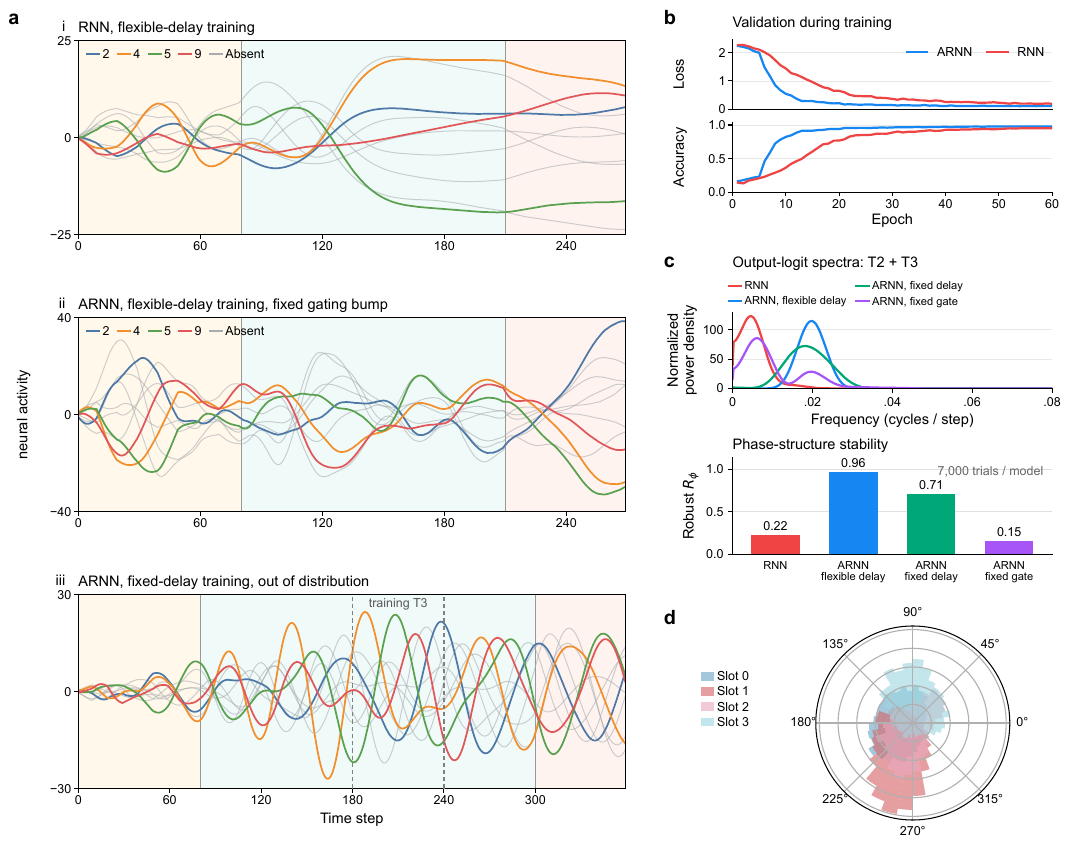}
\caption{\textbf{Adaptation, temporal gating, and delay variability shape phase coding.} (a) Activities of content readout neurons from a non-adaptive RNN (i), an \arnn{} with a fixed temporal gate (ii), and an \arnn{} trained with a fixed delay but tested with a longer delay (iii). Dashed lines in (iii) mark the trained retrieval interval (steps 180--240); the test cue arrives at step 300. The training-delay reference is shown in \figref{fig:fixed-delay-training}. (b) Validation loss and accuracy during training. (c) Output-logit spectra (top) and robust phase consistency $R_\phi$ (bottom). (d) Overlapped phase distributions for serial positions in the fixed-delay network. Settings and analyses appear in \appref{app:conditions}.}
\label{fig:conditions}
\end{figure}

To better understand the emergence of OPC, we inspect the contributions of three key factors by replacing each of them in turn. The three factors are: (i) adaptation in the neural dynamics, (ii) the flexible temporal gating, and (iii) the variability of delay duration. The results are shown in \figref{fig:conditions}a(i--iii); model configurations and analysis details are provided in \appref{app:conditions}.

\noindent\textbf{Replacing the \arnn{} with a standard RNN.} First, we examine the origin of oscillatory activity by comparing the \arnn{} with a standard RNN without adaptation. The non-adaptive network expresses remembered content through slowly varying trajectories, instead of rhythmic item reactivation (\figref{fig:conditions}a(i)). Its spectrum is dominated by low frequencies, whereas the \arnn{} has a prominent oscillatory peak (\figref{fig:conditions}c). The displayed adaptive run also reaches higher final validation accuracy (\figref{fig:conditions}b).

\noindent\textbf{Replacing flexible readout timing with a fixed decision window.} Second, we train an adaptive network to base its decision primarily on activity within a fixed time window near the end of retrieval. We implement this window by replacing the gating head with a Gaussian temporal gate that scales the content-head outputs. This gate peaks at the end of retrieval, and its width is learned jointly with the network. Every trial uses the same temporal gate, regardless of which digits were presented or which one is cued (\appref{app:conditions}). After training, the content-readout activities still oscillate alongside slower changes (\figref{fig:conditions}a(ii),c), but the phase differences between item readouts vary more across trials than in the main \arnn{} (\figref{fig:conditions}c, bottom).

\noindent\textbf{Replacing variable delays with a fixed delay.} Third, we train the network with a fixed maintenance delay. 
At the training delay, oscillatory activity persists (\figref{fig:fixed-delay-training}), but the position-phase distributions overlap substantially (\figref{fig:conditions}d). When the delay is extended, the cue arrives later, 
while prominent item-dependent activity remains near the originally trained retrieval period (\figref{fig:conditions}a(iii)). This temporal organization is tied to the expected retrieval time.

Robust $R_\phi$ measures the stability of oscillatory phase coding through cross-trial consistency in phase differences between item readouts at different ordinal positions, evaluated over temporal windows (\appref{app:conditions}). Only the variable-delay \arnn{} exceeds $0.9$ (\figref{fig:conditions}c, bottom). Together with the trajectories and phase distributions, this result shows that stable oscillatory phase coding emerges in the variable-delay \arnn{}, whereas the comparison networks do not form stable phase representations of ordinal position, even when their activity oscillates.

\subsection{Content and time cells in trained RNNs}
\label{sec:cells}
\begin{figure}[t]
\centering
\includegraphics[width=\linewidth]{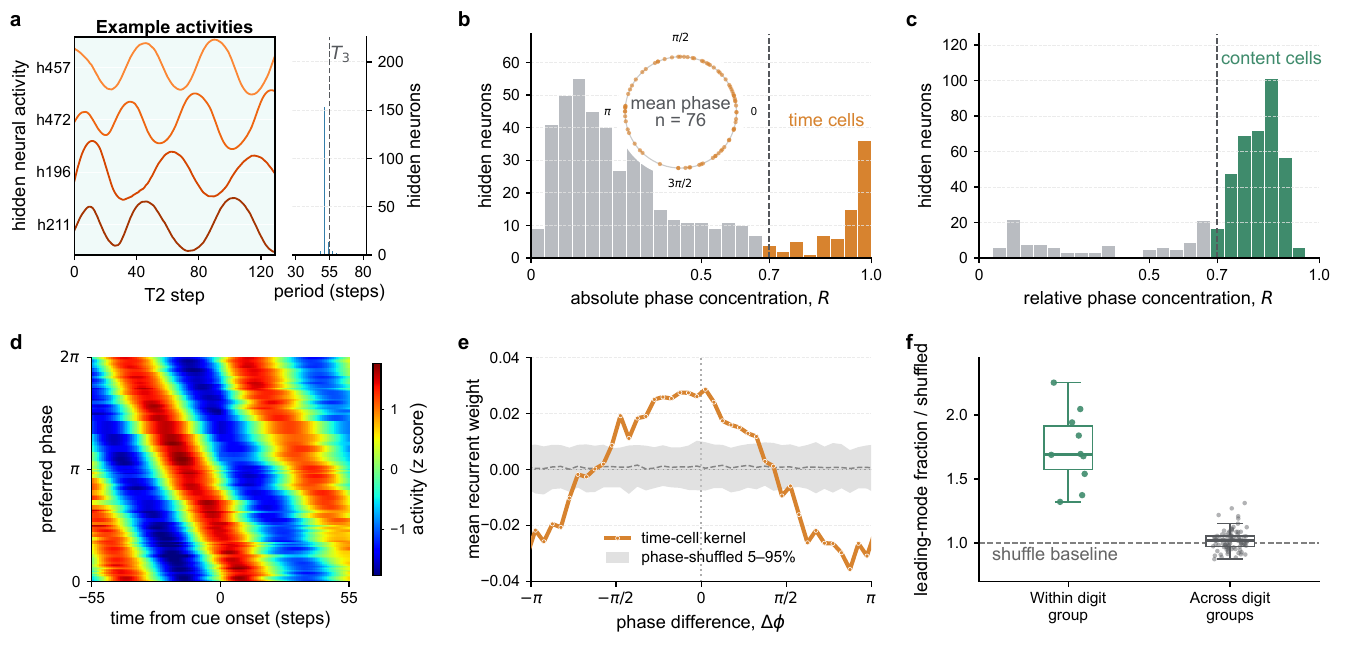}
\caption{\textbf{Time and content cells in the trained A-RNN.} (a) Example neural activity traces (left) and the estimated oscillation periods (right); the dashed line marks $T_3$. (b,c) Absolute (b) and content-relative (c) phase concentration across trials. Colored bars mark neurons meeting the classification criteria; the inset in (b) shows the preferred phases of time-cell candidates. (d) Time-cell activity averaged across trials and z-scored within each neuron. Rows are ordered by preferred phase, revealing a traveling wave across cue onset at time zero. (e) Recurrent connectivity kernel of time cells on the phase ring, showing an approximately cosine shape. (f) Leading-mode fraction within and across digit-selective content-cell groups, normalized by the mean shuffled fraction. The fraction is the proportion of squared, centered weights captured by the dominant singular mode. Phase references, classification criteria, and connectivity analyses are given in \appref{app:cells}.}
\label{fig:cells}
\end{figure}
We next examined single-neuron activity in the trained \arnn{}. Hidden neurons exhibit oscillatory activity, with periods concentrated around a common value slightly below $T_3$ (\figref{fig:cells}a); their learned adaptation strengths, oscillation amplitudes, and periods are characterized in \appref{app:oscillator-properties}. Because each neuron's activity oscillates with this shared period, we fitted a harmonic at this period to each neuron's activity and summarized the timing of its rhythmic modulation by a phase; the fitting and classification criteria are given in \appref{app:cells}. We then asked how each neuron's phase relates to item identity and serial order across trials (\figref{fig:cells}b,c).

We found two distinct phase patterns. In the first, one group of neurons shows preferred phases that remain consistent across trials and sequence compositions, regardless of which items are presented (\figref{fig:cells}b); these spread over the oscillation cycle (\figref{fig:cells}b, inset). We refer to these neurons as time cells. Because their activity peaks at fixed positions within each cycle, we interpret this population as a cyclic temporal reference for the network, one that does not depend on item content.

In the second pattern, each neuron is phase-locked to one specific digit: when that digit appears at different serial positions, the neuron's phase follows the digit's reactivation (\figref{fig:cells}c). We refer to these neurons as content cells; their phases track the reactivation of a particular item, suggesting that they carry item identity. Grouping content cells by their preferred digits yields ten subpopulations, one for each digit 0--9.

Ordering time cells by preferred phase reveals a recurring traveling wave spanning the maintenance and retrieval stages (\figref{fig:cells}d), whereas content cells shift their absolute phase with the serial position of their preferred digit. These complementary patterns suggest a division of labor: content-cell phases remain locked to the time-cell wave, associating each item with its position through phase. This reference is cyclic and does not encode monotonically elapsed time as in interval-timing studies \citep{macdonald2011}. The organization recurs across independently trained networks (\appref{app:robust-cells}).

To relate activity to structure, we examined recurrent connections within each population. The time-cell kernel is approximately cosine-shaped on the phase ring---positive for nearby phases and negative near opposite phases---and disappears under phase shuffling (\figref{fig:cells}e); content cells show structured connectivity within digit-selective groups (\figref{fig:cells}f). Both connectivity patterns recur across independently trained networks (\appref{app:robust-connectivity}; \figref{fig:multiseed-connectivity}).

\subsection{A neural circuit model of OPC inspired by the trained A-RNN}
\label{sec:constructive}

\begin{figure}[t]
\centering
\includegraphics[width=0.8\linewidth]{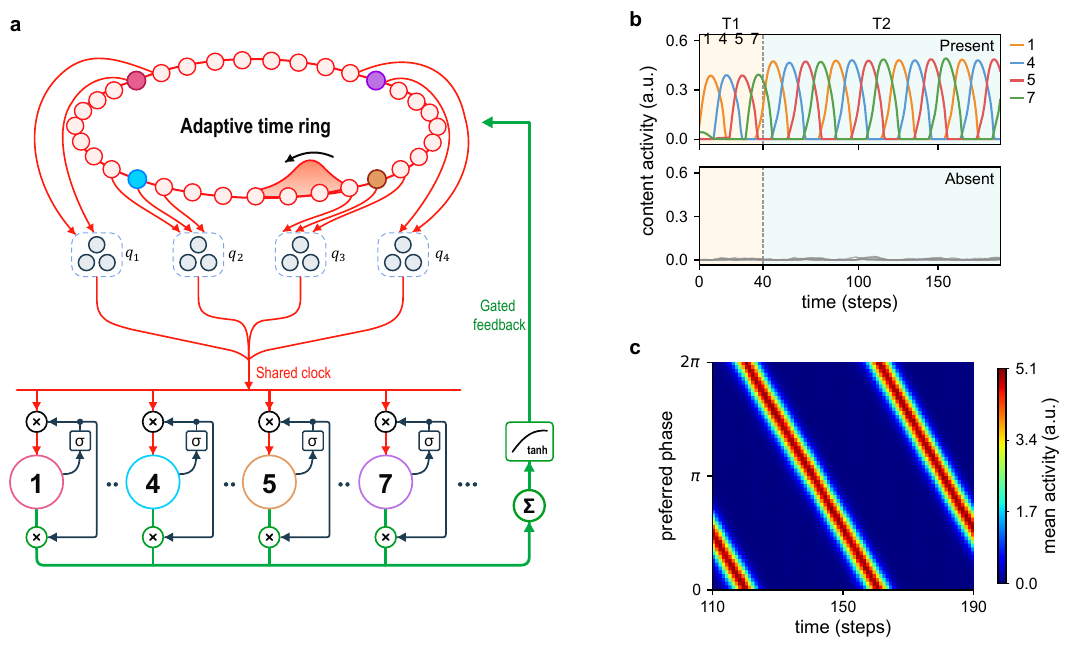}
\caption{\textbf{A hand-designed neural circuit model of OPC.} (a) Circuit model coupling content units to an adaptive time ring through activity-dependent coupling gates and feedback. (b) Content-unit activity for present (top) and absent (bottom) items. (c) Mean time-ring activity across sequences, with units ordered by phase. Model dynamics and analysis details appear in \appref{app:circuit}.}
\label{fig:circuit}
\end{figure}

Guided by these activity and connectivity patterns, we propose a neural circuit model of OPC (\figref{fig:circuit}a) with two reciprocally coupled populations: one content unit per digit-selective content-cell group, representing its population-averaged activity, and a population of time cells forming a ring attractor network with adaptation, which provides a temporal oscillation reference through periodic traveling waves.

Both populations combine recurrent excitation with adaptation. Their synaptic inputs obey
\begin{equation}
\tau_U^C\frac{d\,\mathbf U^C}{dt}=-\mathbf U^C+\tanh\!\left(W_C\mathbf r_C+\lambda\,\mathbf g\odot(W_{T2C}\mathbf r_T)+\mathbf I^C\right)-\mathbf V^C,
\label{eq:circuit-content}
\end{equation}
\begin{equation}
\tau_U^T\frac{d\,\mathbf U^T}{dt}=-\mathbf U^T+W_T\mathbf r_T+\eta\tanh\!\left(W_{C2T}(\mathbf g\odot\mathbf r_C)\right)+\mathbf I^T-\mathbf V^T,
\label{eq:circuit-time}
\end{equation}
with the adaptation of each population $P\in\{C,T\}$ following $\tau_V^P\,\mathrm{d}\mathbf V^P/\mathrm{d}t=-\mathbf V^P+m^P\mathbf U^P$. Here $\mathbf U$ denotes synaptic input, $\mathbf V$ adaptation, $\mathbf r_T$ the normalized firing rates of the ring, $\mathbf r_C$ the firing rates of the content units, $\mathbf I^C$ and $\mathbf I^T$ the external inputs to the content units and the time cells, and $\mathbf g$ an activity-dependent gate (specified below); $\odot$ is elementwise multiplication, and $\lambda$ and $\eta$ set the two coupling strengths.

We first describe the time cells. Their recurrent connectivity $W_T$ is a bell-shaped Gaussian kernel, and their activation function is a divisive normalization that provides global inhibition,
\begin{equation}
r_{T,j}=\frac{(U_j^T)^2}{1+\kappa\rho\sum_\ell(U_\ell^T)^2},
\label{eq:circuit-ring-rate}
\end{equation}
where $\rho$ is the effective neural density and $\kappa$ the normalization strength. This local excitation with global inhibition forms a continuous attractor neural network (CANN) with localized bump attractor states; under adaptation, its intrinsic dynamics turn the bump into a periodic traveling wave \citep{mi2014}. The external input $\mathbf I^T$---a localized bump that sweeps around the ring during encoding---sets the initial moving direction of this wave. We set the global inhibition $\kappa$ especially strong, so that the input from the content units acts as a switch for the wave. With all content units silent, the time cells cannot hold a bump and the ring stays at rest; with content units active, the uniform excitatory input through $W_{C2T}$---an all-to-all uniform excitatory connection---sustains periodic traveling waves, and each time cell activates periodically with a stable phase (\figref{fig:circuit}c). Parameters and mathematical details are included in \appref{app:circuit}.

Within the content population, $W_C=w_{\mathrm{self}}I+J^C$ consists of two parts: one is the self-excitation $w_{\mathrm{self}}I$ with $I$ the identity matrix, and the other is relatively weak random connections between units $J^C$; the self-excitation together with adaptation gives each content unit intrinsic oscillatory dynamics. The external input $\mathbf I^C$ delivers a positive pulse to the unit of each presented digit in succession. Across trials, the memorized digits differ, so the time cells must bind to different content units; we therefore let each content unit's own activity control a gate on the ring input it receives, $g_i=G(r_{C,i})$. This gating is reminiscent of the NMDA receptor, which passes synaptic input only when the postsynaptic neuron is depolarized \citep{nowak1984}; $G$ is a smooth threshold, $G(r)=\max\{s(r-0.25)-s(-0.25),0\}/[1-s(-0.25)]$ with $s(x)=[1+\tanh(10x)]/2$. Only content units activated by the external pulses pass this gate and receive the term $\lambda g_iS$ of Eq.~\eqref{eq:circuit-content}---the time cells' input, where $S$ is the pooled activity of the time cells encoding the four serial-order phases. This gated binding absorbs the small noise-induced mismatches between the oscillation periods of the two populations, allowing a stable phase relationship to persist long-term; the content units thus exhibit stable oscillatory phase coding (\figref{fig:circuit}b). Both external inputs end at maintenance onset, leaving weak Gaussian input noise (\appref{app:circuit}).

The coupled circuit reproduces the OPC signature of the trained \arnn{}: across all ordered four-item sequences, amplitudes distinguish present from absent content units, and phases cluster by serial position while overlapping when grouped by digit identity (\appref{app:circuit-opc}; \figref{fig:circuit-support}a--c). The model thus translates the content--time organization of the trained \arnn{} into a candidate circuit mechanism for biological OPC.

\section{Discussion}
\label{sec:discussion}
\subsection{Contributions}
From a computational neuroscience perspective, our work provides a computational account of how OPC can develop to support sequence working memory. OPC emerges through task learning without an externally imposed rhythm or phase supervision, linking its formation to neural adaptation and the demands of flexible memory retrieval. The content and time cells identified through activity and connectivity analyses inspire a new neural circuit model of biological OPC. This model distills the learned organization into interactions between adaptive content and time units, offering a candidate implementation of phase-coded memory. The study also generates testable predictions: changing an item's serial order in a sequence will shift the phase of the corresponding content cells while preserving the phases of time cells. Our study may also inspire the development of AI models for realizing flexible working memory.

\subsection{Related works}
\paragraph{Theory of OPC.}
The theta--gamma nested theory proposed that sequentially memorized items occupy distinct phases of a shared rhythm \citep{lisman1995,lisman2005}, consistent with phase-dependent item representations in human intracranial recordings \citep{bahramisharif2018}. Adaptation-based network models explained how intrinsic dynamics can generate phase sequences in spatial navigation \citep{chu2024,ji2025}. Our work demonstrated for the first time how OPC emerges in A-RNNs through learning a flexible working memory task, providing insight into the implementation of OPC in neural systems.

\paragraph{Training RNNs for working memory.}
Task-trained RNNs have been used to study attractor dynamics \citep{Sussillo2013}, memory maintenance by short-term synaptic plasticity \citep{kozachkov2022}, and dynamic population coding \citep{stroud2024}. RNNs can also develop phase-coded memories when supplied with a reference oscillation and trained on phase-specific outputs \citep{pals2024}. In contrast, we showed that training RNNs 
with additional adaptive dynamics and flexible computational demand
can give rise to OPC without oscillatory input or phase supervision.

\paragraph{Adaptation in AI models.}
Adaptation has been incorporated into spiking neural networks to support information processing over long time scales \citep{bellec2018,salaj2021} and into neuromorphic systems for efficient computation \citep{ganguly2024}. Here, we explored how adaptation affects the learned strategy in RNNs for sequence working memory, and demonstrated that it plays a critical role in generating OPC.

\subsection{Limitations and future directions}
The predicted content and time cells remain to be verified in biological systems, and we have suggested an experiment
to test this prediction (see \appref{app:biological-test}). Our simplified neural circuit model implements dynamic binding through prescribed activity-dependent coupling between content units and the time ring. It illustrates how this interaction can sustain phase-coded memories, but leaves open how biological circuits realize the coupling. Further modeling studies should investigate the cellular and synaptic mechanisms for this binding, and determine whether it contributes to sustaining phase coding in spiking neural networks.

\label{maintextend}
\clearpage
\section*{AI use statement}
In this work, we used generative AI tools to assist with drafting and polishing portions of the manuscript, including improving clarity, grammar, and organization. We also used generative AI tools to support literature retrieval and discovery of potentially relevant prior work, as well as code development and debugging. All AI-suggested references were independently checked against the original sources, and all AI-assisted mathematical arguments, code, experimental procedures, and results were reviewed and verified by the authors. We made all research decisions and take full responsibility for the final content of this work, including its text, claims, analyses, and results.

\section*{Reproducibility statement}
\appref{app:training} specifies the task, model, and stored training configuration. \appref{app:phase}--\ref{app:ablation-multiseed} define the analysis windows, estimators, cell classes, recurrent connectivity, and ablation comparisons; \appref{app:circuit} gives the hand-designed circuit and numerical integration. Statistical units and checkpoint differences are stated with the corresponding results. We will publicly release the project code repository and trained model checkpoints upon acceptance of the paper.

\section*{Ethics statement}
This study analyzes computational models using MNIST images and synthetic circuit inputs. It includes no newly collected human or animal data. The biological interpretation is a hypothesis about neural computation based on artificial networks and the cited experimental literature.

\bibliography{references}
\bibliographystyle{iclr2027_conference}
\clearpage
\appendix
\section{Task, architecture, and training details}
\label{app:training}
The dataset samples $N$ distinct digit identities without replacement, permutes their order, and draws one MNIST image for each identity \citep{lecun1998}. The query position is uniform over the $N$ positions. For trial $\ell$ and pixel $p$, independent noise $\epsilon^{\mathrm{img}}_{\ell,p}\sim\mathcal N(0,\sigma_{\mathrm{img}}^2)$, with $\sigma_{\mathrm{img}}=0.05$, is added before clipping to $[0,1]$. Pixel noise is drawn once when the trial image is generated; the encoded image is reused across its presentation steps. Training images use the MNIST training split; validation and the subsequent figure analyses use its test split. Unless stated otherwise, each analysis draws an independently generated batch of at least 7,000 trials from this split (Table~\ref{tab:training}), and the analyzed maintenance delay is analysis-dependent: fixed delays between 100 and 220 steps in the delay-specific analyses, or delays drawn from the variable training range. The main model's accuracy on these test batches is approximately $97.9\%$.

\subsection{Encoder and readout implementation}
\label{app:readout}

The main model uses $N=4$ digits, each represented by a $28\times28$ grayscale image. The same CNN is applied independently to each digit. Its layer sequence is listed below; output shapes use channel $\times$ height $\times$ width, with the batch dimension omitted.

\begin{center}
\setlength{\tabcolsep}{6pt}
\renewcommand{\arraystretch}{1.10}
\begin{tabular}{@{}lll@{}}
\toprule
Layer & Configuration & Output shape\\
\midrule
Input & One grayscale digit & $1\times28\times28$\\
Conv2D & $5\times5$, stride 2, padding 2 & $16\times14\times14$\\
BatchNorm2D + ReLU & 16 channels & $16\times14\times14$\\
Dropout2D & Drop probability $0.2$ (training only) & $16\times14\times14$\\
Conv2D & $3\times3$, stride 1, padding 1 & $16\times14\times14$\\
BatchNorm2D + ReLU & 16 channels & $16\times14\times14$\\
Adaptive average pooling & Spatial output $4\times4$ & $16\times4\times4$\\
Flatten & Per-digit feature count & $256$\\
\bottomrule
\end{tabular}
\end{center}

The four encoded blocks are concatenated along the width dimension into a $16\times4\times16$ feature map and flattened to 1,024 features. During a digit's ten-step presentation, only its feature block is active; all other blocks are zero. There are two presentation cycles and no inter-item blank steps. The CNN features are computed once per trial and reused throughout the encoding sequence. A separate 16-dimensional cue stream contains noise before cue onset and the rectified learned embedding afterward. A maintenance delay is drawn uniformly from the integers 50--150 and shared within each minibatch. All durations are simulation steps, without an assigned physical time step.

During maintenance and retrieval, sample features are replaced by $\boldsymbol\epsilon^{(s)}_{\ell,t}\in\mathbb R^{1024}$; before cue onset, cue features are replaced by $\boldsymbol\epsilon^{(c)}_{\ell,t}\in\mathbb R^{16}$. Their scalar entries satisfy
\begin{equation}
\epsilon^{(s)}_{\ell,t,k},\ \epsilon^{(c)}_{\ell,t,j}
\overset{\mathrm{i.i.d.}}{\sim}\mathcal N(0,\sigma_{\mathrm{feat}}^2),
\qquad \sigma_{\mathrm{feat}}=0.05.
\label{eq:input-noise}
\end{equation}
These feature-space noises are sampled independently at every simulation step, across trials and feature coordinates, and independently between the sample and cue streams.

We set $\alpha_h=0.1$ and $\alpha_\theta=0.01$, guided by the separation of neural and adaptation timescales in the model of \citet{mi2014}. The smaller adaptation update rate reflects the biologically motivated assumption that spike-frequency adaptation evolves more slowly than neural activity.

Equations~\eqref{eq:drive}--\eqref{eq:adaptation} follow the implemented update order, including subtraction of the previous adaptation state outside the $\alpha_h$-scaled term. Layer normalization has no learned affine parameters. Each adaptation gain is parameterized as $m_i=\operatorname{softplus}(\widetilde m_i)+10^{-6}$, and $\boldsymbol\gamma=\mathbf{1}$. Input connection weights are constrained to be nonnegative by ReLU; recurrent and readout connection weights are unconstrained. Hidden and adaptation states start at zero. The adaptation state is driven by signed $\mathbf h_t$, and the content logits and unnormalized gating scores are linear functions of $\mathbf h_t$. Temporal softmax is applied only to the gate scores over retrieval time.

The CNN operator in Eq.~\eqref{eq:drive} includes the image encoder, sequence-dependent feature masking, and the input projection into the recurrent population. The cue encoder applies a learned embedding, ReLU, and the corresponding input projection. For a one-hot query $\mathbf u_q$, an embedding lookup with matrix $E$ is equivalent to $E\mathbf u_q$, giving the feedforward map $W_c\operatorname{ReLU}(E\mathbf u_q)$ denoted by MLP. During maintenance and retrieval, sample features are replaced by noise before the input projection; cue features are likewise replaced by noise before retrieval. The implemented adaptation feedback has fixed gain $\boldsymbol\gamma=\mathbf 1$, which is absorbed into the subtraction in Eq.~\eqref{eq:activity}.

At each retrieval step, the content and gating heads compute
\begin{equation}
\mathbf p_t=W_{\mathrm{out}}\mathbf h_t+\mathbf b_{\mathrm{out}},
\label{eq:content-head}
\end{equation}
\begin{equation}
g_t=\mathbf w_g^{\mathsf T}\mathbf h_t+b_g.
\label{eq:gate-score}
\end{equation}
The vector $\mathbf p_t$ contains raw digit logits, and $g_t$ is an unnormalized scalar gate score. For the set $\mathcal T_3$ of retrieval time indices, the temporal gating values are
\begin{equation}
\alpha_t=\left[\operatorname{Softmax}\!\left((g_s)_{s\in\mathcal T_3}\right)\right]_t
=\frac{\exp(g_t)}{\sum_{s\in\mathcal T_3}\exp(g_s)}.
\label{eq:temporal-gate}
\end{equation}
Softmax is applied over the complete retrieval window, so the gating values sum to one and Eq.~\eqref{eq:readout} reduces to $\sum_t\alpha_t\mathbf p_t$. The predicted digit is $\arg\max_d\widehat y_d$. The trajectories analyzed in the figures are the raw logits $p_{t,d}$, before any class normalization. The response is computed from the gated content readouts over the complete retrieval stage.

\noindent\textbf{Parameter budgets in RNN--\arnn{} comparisons.}
At a fixed hidden size $H$, adaptation adds the learnable gain vector $\widetilde{\mathbf m}\in\mathbb R^H$, giving $H$ additional trainable parameters, or $O(H)$ overhead relative to the $O(H^2)$ recurrent weight matrix. The adaptation state $\boldsymbol\theta_t$ is a dynamical variable, not an additional set of trainable parameters; its update rate and feedback gain are fixed. We therefore give the non-adaptive RNN a slightly larger hidden population to keep parameter budgets comparable: 502 versus 500 units in \appref{app:conditions}, and $H+5$ versus $H$ units in \appref{app:s1}. These offsets approximately match parameter budgets while giving the non-adaptive RNN the slightly larger budget.

\subsection{Training configuration}
\label{app:optimization}
The model is trained through backpropagation through time using AdamW \citep{loshchilov2019}, with checkpoint selection based on validation accuracy. The hidden population has no prescribed content/time partition or ring connectivity.

\begin{table}[h]
\centering
\caption{Representative defaults and typical ranges for training the recurrent models. Experiment-specific network sizes, task timings, and training budgets are given in the corresponding appendix sections.}
\label{tab:training}
\begin{tabular}{ll}
\toprule
Quantity & Default or typical range\\
\midrule
Hidden size $H$; cue embedding dimension & 100--500; 16\\
Update rates $\alpha_h$; $\alpha_\theta$ & $0.1$; $0.01$\\
Initial adaptation gain $\mathbf m$; feedback gain $\boldsymbol\gamma$ & $[0.9,1.1]$; $1.0$\\
Encoding $T_1$ & 10 steps per digit, 2 rounds ($T_1=20N$)\\
Inter-item blank interval & 0 steps\\
Delay $T_2$; retrieval $T_3$ & Integer uniform $[50,150]$; 55--65 steps\\
Pixel-noise standard deviation & $\sigma_{\mathrm{img}}=0.05$ (once per trial)\\
Feature/cue-noise standard deviation & $\sigma_{\mathrm{feat}}=0.05$ (i.i.d. each step; Eq.~\eqref{eq:input-noise})\\
Training/validation trials per epoch & 75,000 / 15,000\\
Test trials per analysis batch & $\geq$7{,}000 unless stated otherwise\\
Training/validation batch size & 64 / 64\\
Training budget & $>60$ epochs\\
Optimizer; gradient norm clipping & AdamW; $1.0$\\
Main learning rate: peak / minimum & $1.5\times10^{-4}$ / $1.5\times10^{-5}$\\
Adaptation learning rate: peak / minimum & $1\times10^{-3}$ / $1\times10^{-5}$\\
Main/adaptation weight decay & $1\times10^{-4}$ / $1\times10^{-4}$\\
Learning-rate scheduler & 5--10 warmup epochs, then cosine decay\\
Pretraining; mixed precision & None; disabled\\
\bottomrule
\end{tabular}
\end{table}

The training loss is
\begin{equation}
\mathcal L=-\log\frac{\exp(\widehat y_{y_{\mathrm{true}}})}{\sum_{d=0}^{9}\exp(\widehat y_d)}.
\end{equation}
Here $y_{\mathrm{true}}$ is the digit identity at the queried sequence position. No auxiliary loss terms are used; in particular, no loss term refers to a phase, period, cell class, or oscillatory waveform.

The hand-designed circuit in \figref{fig:circuit}a--c uses 60 time units, ten content units, and four local receptive fields. Its time-to-content and content-to-time gains are $\lambda\simeq1.07$ and $\eta=0.5$, respectively. Encoding lasts 40 steps, with four sequential ten-step inputs, followed by 150 maintenance steps. The circuit is fixed across all 5,040 ordered sequences of four distinct digits. The remaining circuit constants, noise settings, and numerical integration rules are specified with their defining equations in \appref{app:circuit}; analysis settings for the trained networks are given in Appendices~\ref{app:phase}--\ref{app:ablation-multiseed}.

\paragraph{Scope of the evaluation and comparisons.}
Our evaluation tests newly sampled trials from the same task distribution and does not address generalization to disjoint held-out digit combinations or repeated identities. The ablations in \appref{app:conditions} and \appref{app:ablation-multiseed} examine the contributions of neural adaptation, state-dependent temporal gating, and variable-delay training to phase coding. Their network sizes, temporal protocols, and training configurations are specified in the corresponding sections. The recurrent-connectivity analysis in \appref{app:connectivity} is performed on the main trained network, while the underlying time/content organization recurs across independently trained networks (\appref{app:robust-cells}).

\section{Amplitude, phase, and gate alignment}
\label{app:phase}
For a trace $x(t)$ over a window of $L$ samples, temporal demeaning precedes a least-squares fit to a fixed-frequency basis,
\begin{equation}
x(t)-\overline x\simeq \beta_1\sin(2\pi t/P)+\beta_2\cos(2\pi t/P),\qquad
A=\sqrt{\beta_1^2+\beta_2^2},\quad \phi=\operatorname{atan2}(\beta_2,\beta_1)\bmod 2\pi.
\label{eq:harmonic}
\end{equation}
\subsection{Amplitude separation and retrieval performance}
\label{app:amplitude-performance}
For the main \arnn{}, the reference period is $P=55$ steps. The amplitude distributions in \figref{fig:code}b use the final 55 pre-cue steps, first averaging over the four present or six absent channels within each trial, so that each group contains one value per evaluation trial, analyzed at $T_2=130$. Mean trial-averaged amplitudes are $20.71$ for present digits and $5.48$ for absent digits. The density plots use separate Scott-bandwidth kernel estimates; their dashed lines indicate unsmoothed means. These amplitudes are in raw-logit units.

\begin{figure}[!t]
\centering
\includegraphics[width=0.90\linewidth]{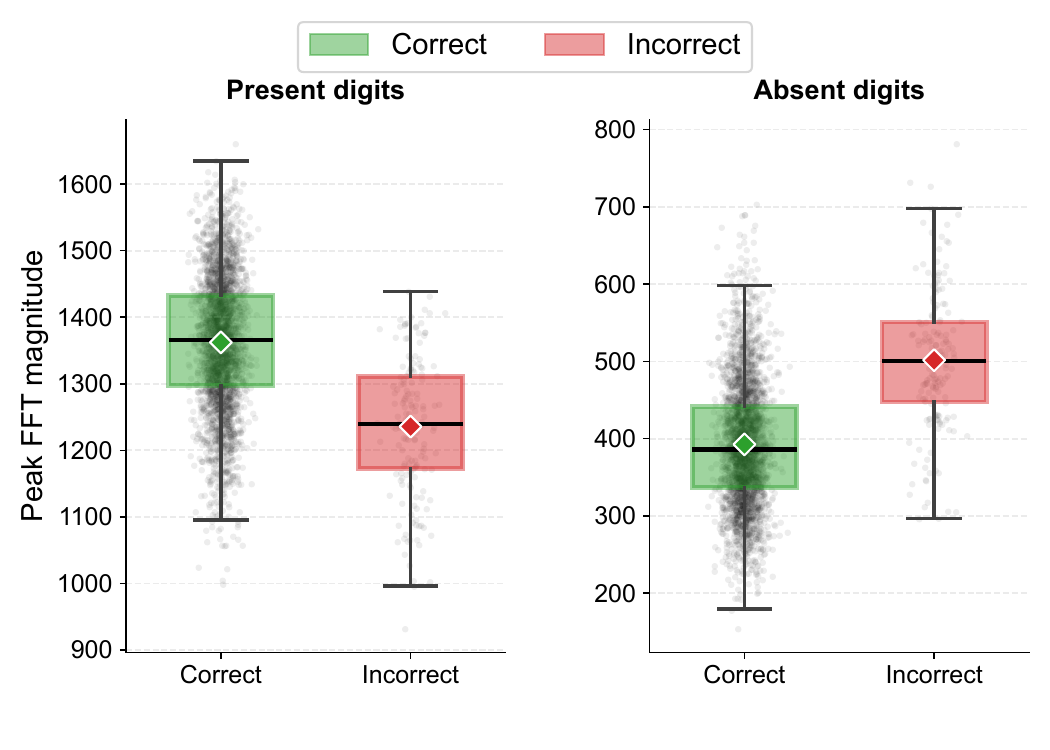}
\caption{\textbf{Retrieval errors accompany weaker present-digit and stronger absent-digit oscillations.} Per-trial mean peak FFT magnitudes over $T_2+T_3$ for present (left) and absent (right) digits, split into correct (green; $n=7{,}834$) and incorrect (red; $n=166$) trials. Vertical scales differ between panels. Gray points show sampled trials, diamonds mark full-group means, and boxes show medians and interquartile ranges; whiskers reach the most extreme values within 1.5 interquartile ranges.}
\label{fig:amplitude-performance}
\end{figure}

Figure~\ref{fig:amplitude-performance} compares peak Fourier magnitudes between correct and incorrect retrieval over the complete maintenance and retrieval interval. For each digit channel $d$ in trial $\ell$, let $p_{\ell,t,d}$ denote its raw logit after encoding. With $T_2=130$, $T_3=55$, and $L=T_2+T_3=185$, we compute
\begin{equation}
M_{\ell,d}=\max_{1\leq k<\lfloor L/2\rfloor}
\left|\sum_{t=0}^{L-1}p_{\ell,t,d}\exp\!\left(-\frac{2\pi\mathrm{i}kt}{L}\right)\right|.
\label{eq:outcome-fft}
\end{equation}
The DC bin is excluded. No taper, zero-padding, squaring, or length normalization is applied, so $M_{\ell,d}$ is an unnormalized FFT magnitude, not spectral power or the fixed-frequency harmonic amplitude used in \figref{fig:code}b. Magnitudes are averaged over the four present or six absent digit channels within each trial, then grouped by retrieval outcome. The analysis uses the same checkpoint; 7,834 trials are correct and 166 incorrect. Mean present-digit magnitudes are 1,362 and 1,236 for correct and incorrect trials, respectively; mean absent-digit magnitudes are 393 and 502. Thus, errors are associated with weaker present-digit and stronger absent-digit oscillatory signals (\figref{fig:amplitude-performance}). Box statistics and means use all trials.

\subsection{Ordinal phase distributions}
\label{app:ordinal-phase}

The phase histograms in \figref{fig:code}c,d use an independently generated evaluation batch at $T_2=130$, and fit the final 55 maintenance steps together with all 55 retrieval steps. Fits with $R^2<0.5$ are excluded, leaving over 99.9\% of trials valid at each position. A single rotation places the circular mean of position 0 at $\pi$ and is applied to every observation. The circular means are approximately $180.0^\circ$, $89.4^\circ$, $356.3^\circ$, and $273.1^\circ$, with resultant lengths between $0.977$ and $0.982$. This decreasing order follows the sine/cosine phase convention. The 36-bin histograms are normalized densities. Digit-based histograms regroup the same phases by identity, without per-digit or per-trial alignment. The shared phase sectors across digit identities provide evidence that ordinal phase is separable from item identity.

\subsection{Temporal gate alignment}
\label{app:gate-alignment}
For gate alignment, an independent evaluation batch with variable $T_2\sim\mathrm{Uniform}\{50,\ldots,150\}$ is used. Let $u_\ell$ be the argmax time of the ground-truth digit logit during retrieval of trial $\ell$ and $v_\ell$ the gate argmax, both indexed from 0 to 54. The circular residual and plotted gate coordinate are
\begin{equation}
\delta_\ell=\bigl((v_\ell-u_\ell+P/2)\bmod P\bigr)-P/2,
\qquad \widetilde v_\ell=u_\ell+\delta_\ell,
\qquad P=55.
\end{equation}
Statistics and the displayed scatter use the correctly classified trials. The phase-locking value is $|\langle\exp(2\pi\mathrm{i}\delta_\ell/P)\rangle_\ell|$, which equals $0.964$; $88.64\%$ of circular absolute errors are within three steps. The plot can place $\widetilde v_\ell$ outside the retrieval window through unwrapping, although the actual gate peak always lies within it. Raw noncircular absolute error is $6.19$ steps, versus circular absolute error $1.71$ steps. These definitions distinguish phase-equivalent timing from equality of the original peak indices.

\section{Period estimation and hidden-neuron classification}
\label{app:cells}
The period distribution in \figref{fig:cells}a uses the last 55 pre-cue steps of the evaluation trials. For each trial and neuron, a demeaned, zero-padded 2,048-point real FFT yields a power spectrum with the DC component removed. Spectra are normalized within each trial/neuron and then averaged across trials for each neuron. The spectral peak is refined by local log-parabolic interpolation and converted to period. Zero-padding supplies an interpolation grid without increasing the native frequency resolution of the 55-step window. The four displayed trajectories are chosen at nearly equally spaced phases among neurons whose single-example harmonic fit has $R^2\geq0.9$.

Across the 500 neurons, the median estimated period is $51.58$ steps and the interquartile range is $51.07$--$52.01$ steps. These spectral estimates are distinct from the fixed reference period used for harmonic phase fitting.

Classification uses a fixed-period harmonic fit over the final 55 pre-cue steps. A trial/neuron phase enters circular statistics when its fit has $R^2\geq0.5$; a neuron is eligible for classification when its mean fit $R^2\geq0.8$. For neuron $i$, let $\mathcal N_i$ collect the trials whose phases pass the per-trial criterion, with $N_i=\lvert\mathcal N_i\rvert$. For eligible phases $\phi_{\ell i}$, $\ell\in\mathcal N_i$, the absolute concentration is
\begin{equation}
R_i^{\mathrm{abs}}=\left|\frac{1}{N_i}\sum_{\ell\in\mathcal N_i} e^{\mathrm{i}\phi_{\ell i}}\right|.
\end{equation}
Let $s(\ell,d)$ denote the position of digit $d$ in trial $\ell$, and let $\psi_s$ be the fixed output-derived preferred phase of position $s$. Let $\mathcal N_{i,d}\subseteq\mathcal N_i$ collect the valid trials containing digit $d$, with $N_{i,d}=\lvert\mathcal N_{i,d}\rvert$. The content-relative score is
\begin{equation}
R_{i,d}^{\mathrm{rel}}=
\left|\frac{1}{N_{i,d}}\sum_{\ell\in\mathcal N_{i,d}}
 e^{\mathrm{i}(\phi_{\ell i}-\psi_{s(\ell,d)})}\right|,
\qquad R_i^{\mathrm{content}}=\max_d R_{i,d}^{\mathrm{rel}}.
\label{eq:cellphase}
\end{equation}
This uses a fixed slot-phase reference, not a trial-by-trial subtraction of the preferred digit's instantaneous readout phase. The reference phases come from the pre-cue slot reference used for classification and lie within $2^\circ$ of the display means in \figref{fig:code}c.

Time candidates meet $R_i^{\mathrm{abs}}\geq0.7$, and content cells meet $R_i^{\mathrm{content}}\geq0.7$. There are 76 time-cell candidates and 370 content-cell candidates, including two overlaps. Content assignment takes priority for overlaps, yielding 74 time cells, 370 content cells, and 56 undefined cells. The 370 content neurons are uniquely assigned to the digit maximizing their relative concentration; the group sizes for digits 0--9 are $43,42,36,29,39,36,41,43,28,33$. The time-cell heatmap uses all 74 exclusive time cells, sorted by their absolute preferred phase. Raw activity is averaged across trials over the interval from 55 steps before to 55 steps after cue onset, then z-scored per neuron across that interval. Its color values are not firing rates.

\subsection{Recurrent connectivity of time and content cells}
\label{app:connectivity}
The connectivity analysis uses the recurrent weight matrix of the main trained \arnn{} and the phase-defined cell groups described above. All connections are directed, and self-connections are excluded. The time-cell analysis includes the 74 exclusive time cells, giving 5,402 connections. For a connection from neuron $i$ to neuron $j$, its phase difference is $\Delta\phi_{ij}=(\phi_j-\phi_i+\pi)\bmod 2\pi-\pi$. We average signed recurrent weights in 40 equal bins over $[-\pi,\pi)$ (\figref{fig:cells}e). For the control, we permute preferred phases among the same neurons 200 times while keeping the weights fixed; the dashed curve and band summarize the mean and pointwise 5th--95th percentiles of the resulting kernels.

The content-cell analysis uses all 370 content cells, assigned to digit groups by their preferred item in the phase analysis. Group sizes for digits 0--9 are $43,42,36,29,39,36,41,43,28,33$. We extract ten within-group connection blocks and 90 directed cross-group blocks from the recurrent matrix. The within-group blocks contain 13,600 valid connections in total, and the cross-group blocks contain 122,930. Signed weights are retained. We subtract each block's mean over valid connections and set excluded self-connections to zero in the centered matrix $B$.

To quantify coherent organization within a connection block, we compute its singular values $\sigma_k$ and the fraction
\begin{equation}
f_1(B)=\frac{\sigma_1(B)^2}{\sum_k\sigma_k(B)^2}
      =\frac{\sigma_1(B)^2}{\lVert B\rVert_F^2}.
\end{equation}
The numerator is the sum of squared weights captured by the best rank-one approximation $\sigma_1\mathbf u_1\mathbf v_1^{\mathsf T}$, which describes a shared connection pattern with neuron-specific source and target contributions. The denominator is the sum of squared centered weights in the full block. This fraction measures how much of the block can be described by its strongest common pattern.

For each block, we independently permute the valid signed weights 200 times, preserving the block dimensions, full weight distribution, and excluded diagonal positions. We normalize the observed fraction by its mean across these shuffled matrices:
\begin{equation}
\mathcal S(B)=\frac{f_1(B)}{\langle f_1(B^{\mathrm{shuffle}})\rangle}.
\end{equation}
A value of one corresponds to the shuffled baseline; larger values indicate that a coherent pattern accounts for more of the connection matrix than expected from randomly arranging the same weights. The shuffled normalization accounts for differences in block size and weight variability. Mean $\mathcal S$ is $1.738$ for within-group blocks and $1.023$ for cross-group blocks; the corresponding medians are $1.692$ and $1.019$. The raw cross-group weights have mean $1.6\times10^{-4}$ and standard deviation $0.048$. \figref{fig:cells}f shows every block, with a box spanning the interquartile range, a median line, and whiskers extending to the most extreme values within 1.5 interquartile ranges; horizontal jitter separates points. Values close to baseline indicate little excess structure under this measure. Sensitivity to the cell-classification thresholds is evaluated in \appref{app:threshold-sensitivity}.

\subsection{Robust identification of time and content cells across thresholds}
\label{app:threshold-sensitivity}
\begin{figure}[!t]
\centering
\includegraphics[width=\linewidth]{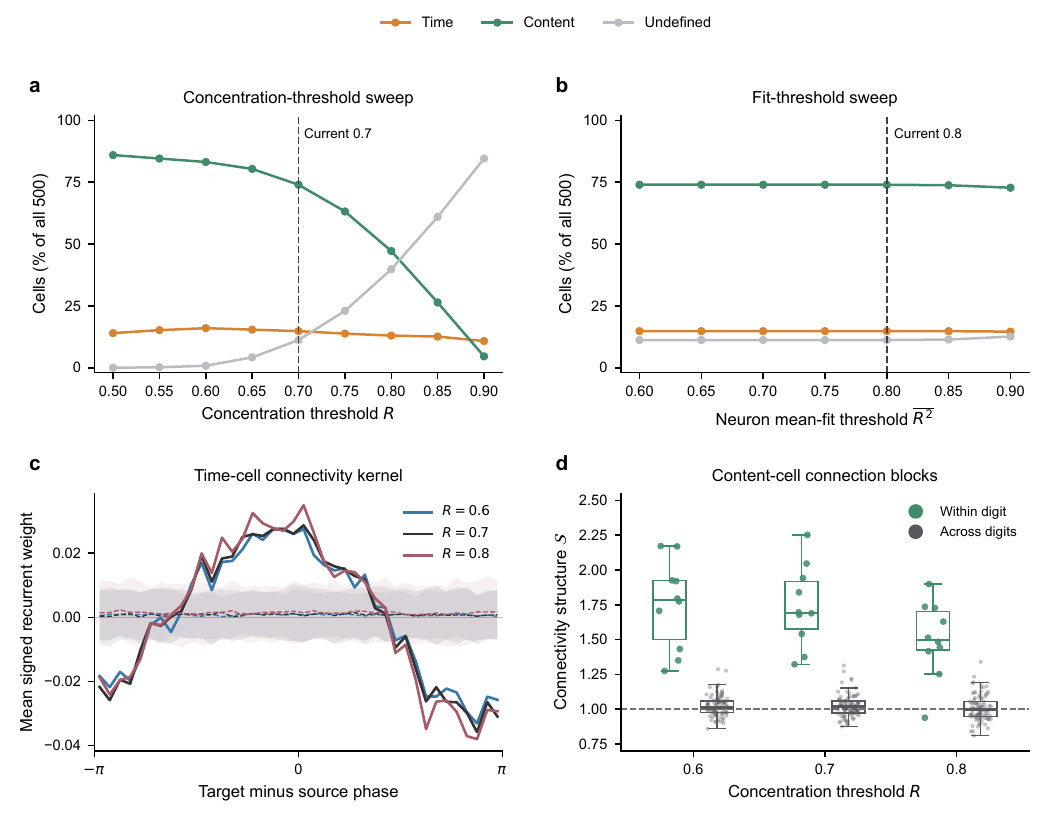}
\caption{\textbf{Time- and content-cell populations remain identifiable across classification thresholds.} (a) Cell fractions as the concentration threshold $R$ varies, with neuron mean-fit qualification fixed at $\overline{R^2}\geq0.8$. (b) Cell fractions as the mean-fit threshold varies, with concentration thresholds fixed at $R\geq0.7$. Dashed lines mark the current settings. (c) Time-cell recurrent kernels for concentration thresholds 0.6, 0.7, and 0.8, with phase-shuffled means and pointwise 5th--95th percentile bands. (d) Within-digit and across-digit content-cell connectivity structure at the same thresholds; points represent directed connection blocks and the dashed line is the shuffled baseline.}
\label{fig:threshold-sensitivity}
\end{figure}

Time- and content-cell populations remain identifiable across the tested classification thresholds. We evaluate their membership and recurrent organization while holding the main checkpoint, phase fits, and slot references fixed. The common concentration threshold for absolute and content-relative phases is scanned from 0.50 to 0.90 in steps of 0.05. The neuron-level mean-fit qualification threshold is scanned from 0.60 to 0.90 in the same increments. The per-trial phase-fit threshold remains $R^2\geq0.5$, and content assignment retains priority when the two criteria overlap. The full scan contains 63 threshold pairs. \figref{fig:threshold-sensitivity}a,b shows one-dimensional slices through the current setting, $R=0.7$ and neuron mean $R^2=0.8$. All fractions use the complete 500-neuron population as denominator.

At $R=0.7$, classification is identical throughout the tested neuron mean-$R^2$ range 0.60--0.80: 74 time cells, 370 content cells, and 56 undefined cells. The lowest neuron mean fit is $0.812$, so all neurons remain eligible over that interval. Raising the qualification threshold to 0.90 gives 73, 364, and 63 cells, respectively. Cell counts are more sensitive to the concentration threshold. With neuron mean $R^2\geq0.8$, the time/content/undefined counts are 80/416/4 at $R=0.6$, 74/370/56 at $R=0.7$, and 65/236/199 at $R=0.8$. Across these three thresholds, 65 neurons are consistently identified as time cells and 236 as content cells; these shared populations contain every neuron selected at the strictest threshold. All ten digit-selective content groups remain represented. The thresholds therefore change group size while preserving a common core of both cell populations.

For each of these three concentration thresholds, we recompute the analyses in \appref{app:connectivity} from the reassigned groups. The time-cell kernel uses directed non-self connections, 40 phase-difference bins, and 200 permutations of preferred phases among the selected neurons. Its shape is summarized by an equal-bin least-squares fit $K(\Delta\phi)=c_0+c_c\cos\Delta\phi+c_s\sin\Delta\phi$. At thresholds 0.6, 0.7, and 0.8, this first harmonic explains $96.33\%$, $97.12\%$, and $95.79\%$ of the across-bin variance. The two alternative-threshold kernels correlate with the current kernel at $0.995$ and $0.993$. The positive near-phase and negative opposite-phase organization is retained.

The content-cell analysis assigns each selected neuron to its highest-concentration digit and evaluates ten within-digit and 90 directed across-digit blocks at each threshold. Each block retains signed weights and uses 200 weight permutations for the leading-mode normalization in \appref{app:connectivity}. Mean within-digit structure is $1.752$, $1.738$, and $1.503$ at thresholds 0.6, 0.7, and 0.8; the corresponding across-digit means are $1.023$, $1.023$, and $1.008$. All ten within-digit blocks exceed their own shuffled 95th percentile at thresholds 0.6 and 0.7, and nine of ten do so at 0.8. The within-digit effect is weaker at the strictest threshold, but the contrast with near-baseline across-digit connectivity remains. Together, the shared cell populations and preserved connectivity patterns support robust identification of time and content groups across the tested thresholds, with threshold-dependent group boundaries.

\section{Trained comparison conditions and robust phase consistency}
\label{app:conditions}
\begin{table}[h]
\centering
\caption{Conditions underlying \figref{fig:conditions}. Each row denotes one trained checkpoint. $T_3^{\mathrm{train}}$ is its training retrieval length, and $Q=T_1+T_2$ is the evaluation cue onset.}
\label{tab:conditions}
\begin{tabular}{lrrrrl}
\toprule
Condition & $N$ & $H$ & $T_1$ & $T_3^{\mathrm{train}}$ & Training delay\\
\midrule
Non-adaptive RNN & 4 & 502 & 80 & 60 & Uniform 50--150\\
\arnn{}, variable delay & 4 & 500 & 80 & 55 & Uniform 50--150\\
\arnn{}, fixed delay & 4 & 500 & 80 & 60 & Fixed 100\\
\arnn{}, fixed gate & 4 & 500 & 80 & 60 & Uniform 50--150\\
\bottomrule
\end{tabular}
\end{table}

The comparisons use the training framework summarized in Table~\ref{tab:training}; their network sizes and temporal protocols are listed above. The RNN hidden-size offset follows the parameter-budget convention in \appref{app:readout}. The displayed learning curves cover the first 60 epochs, with validation accuracies of $97.87\%$ for the main adaptive run and $95.37\%$ for the non-adaptive run at epoch 60.

\begin{figure}[htbp]
\centering
\includegraphics[width=0.82\linewidth]{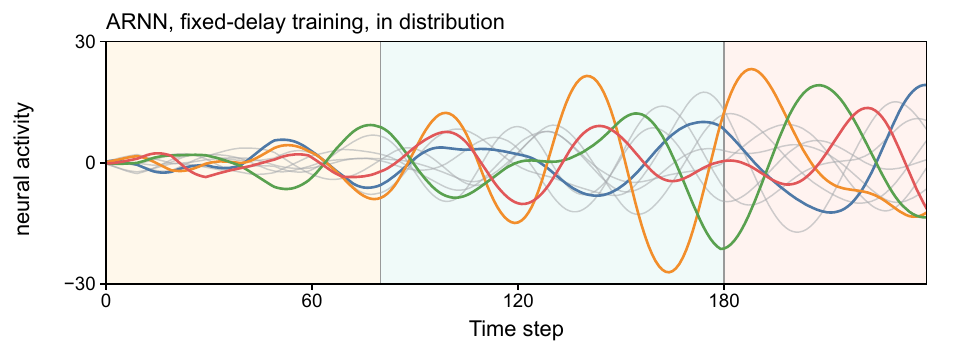}
\caption{\textbf{Content readouts at the fixed training delay.} The fixed-delay network is evaluated at its trained delay, $T_2=100$ steps, providing the reference for \figref{fig:conditions}a(iii). Colored traces represent presented digits and gray traces absent digits; shading marks encoding, delay, and retrieval.}
\label{fig:fixed-delay-training}
\end{figure}

\begin{figure}[t]
\centering
\includegraphics[width=0.68\linewidth]{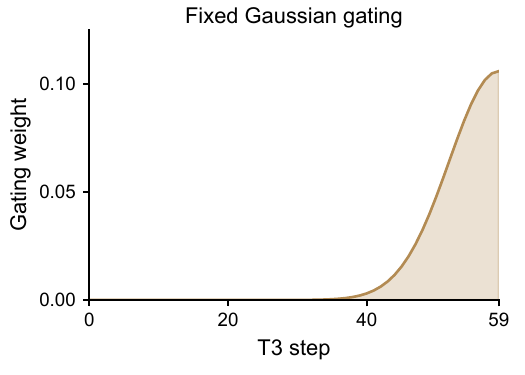}
\caption{\textbf{Fixed temporal gate used in the readout ablation.} The Gaussian gate peaks at the final retrieval step. Its width is learned during training, and every trial uses the same gate regardless of the input digits or cue.}
\label{fig:fixedgate}
\end{figure}

The fixed-gate control replaces Eq.~\eqref{eq:readout}'s state-dependent gating values with
\begin{equation}
\alpha_r^{\mathrm{fixed}}=
\frac{\exp[-(r-59)^2/(2\sigma^2)]}{\sum_{s=0}^{59}\exp[-(s-59)^2/(2\sigma^2)]},
\qquad r=0,\ldots,59.
\end{equation}
The positive width is a globally learned scalar, $\sigma\approx7.14$ in the selected checkpoint, and is fixed across images and cues during inference. The temporal gate is shown in \figref{fig:fixedgate}. The fixed-delay examples (\figref{fig:fixed-delay-training} and \figref{fig:conditions}a(iii)) use $T_2=100$ and $220$, so ``out of distribution'' refers specifically to extension beyond the singleton training delay. The training-duration example uses $T_1=80$, $T_2=100$, and $T_3=60$, with retrieval spanning steps 180--240. Colored traces are the four presented digits (2, 4, 5, and 9), gray traces are absent digits, and shading distinguishes the three task stages. The extended-delay example receives its cue at step 300; its dashed lines retain the training retrieval boundaries.

The spectra in \figref{fig:conditions}c (top) use raw logits over the entire $T_2+T_3$ interval, excluding encoding. Each channel is demeaned; a periodic-Hann periodogram uses one sample per step, a 4,096-point FFT, and density scaling. Raw densities are averaged across the ten output channels and divided by the full $0$--$0.5$ cycles/step integral. No common-mode removal or channel z-score is used for this panel. The four windows contain 190, 185, 160, and 190 steps for the non-adaptive, main adaptive, fixed-delay, and fixed-gate conditions. Each spectrum describes one example.

The fixed-delay phase histogram retains its original estimator: we demean the full 160-step maintenance-plus-retrieval trace, form the FFT analytic signal, and circularly average instantaneous phase over time. The phases are rotated together to place position 0 at $\pi$. This estimator differs from the harmonic regression used for \figref{fig:code}c,d; the robust phase statistic below provides a common definition across all four conditions.

For robust $R_\phi$, each checkpoint is evaluated without filtering errors. Inference retrieval is extended to $4T_3^{\mathrm{train}}$, and the analyzed interval is $[Q-T_3^{\mathrm{train}},Q+2T_3^{\mathrm{train}})$, of length $L=3T_3^{\mathrm{train}}$. The test $T_2$ was 130 for the variable-delay conditions and 100 for the fixed-delay condition. We subtract the mean over all ten logits at each step, select the present-digit channels in serial order, and fit and remove one linear trend over the whole interval. Multiplication by a symmetric Hann window and an rFFT gives coefficients $C_{\ell k\nu}$ for trial $\ell$, serial position $k$, and frequency bin $\nu$. The shared bin $\nu^\ast$ maximizes trial- and position-averaged power over $\nu=2,\ldots,F-2$, where $F=\lfloor L/2\rfloor+1$ is the rFFT bin count.

At the resulting fixed frequency $f_0=\nu^\ast/L$, sliding windows have length $M=\max(4,\operatorname{round}(1/f_0))$ and stride $\max(1,\operatorname{round}(1/(6f_0)))$. The globally detrended trajectories $\widetilde x$ are projected through a local Hann window $h_M$:
\begin{align}
c_{\ell k}(s)&=\sum_{\kappa=0}^{M-1}\widetilde x_{\ell k}(s+\kappa)\,h_M(\kappa)\,e^{-2\pi\mathrm{i}f_0\kappa},\\
R_{kk'}(s)&=\frac{|\sum_{\ell} c_{\ell k}(s)c_{\ell k'}(s)^{*}|}
 {\sum_{\ell}|c_{\ell k}(s)|\,|c_{\ell k'}(s)|+10^{-8}},\\
R_\phi(s)&=\binom{N}{2}^{-1}\sum_{k<k'}R_{kk'}(s),\qquad
R_\phi^{\mathrm{robust}}=\operatorname{percentile}_{20}\{R_\phi(s)\}.
\end{align}
The percentile uses linear interpolation. No new detrending is performed inside a sliding window. Window lengths are 90, 55, 60, and 90 steps, giving 7, 13, 13, and 7 overlapping windows in the table order. Robust $R_\phi$ is $0.964$ for the main \arnn{}, $0.710$ for fixed-delay training, $0.224$ without adaptation, and $0.154$ with fixed gating. In the separate full-window spectral analysis, the main adaptive and non-adaptive examples peak near $0.020$ and $0.005$ cycles per step, respectively. The score summarizes phase-difference consistency; within-trial phase separation is quantified separately (\appref{app:ablation-multiseed}).

\section{Phase structure across ablation runs}
\label{app:ablation-multiseed}
We evaluate the robustness of the ablation outcomes using independently initialized networks with fixed training delays, fixed temporal gates, or no adaptation. The fixed-delay and fixed-gate conditions each use four training seeds (21001--21004), paired across the two conditions. Their sampled configurations $(N,H,T_3)$ are $(3,450,65)$, $(3,500,65)$, $(4,550,65)$, and $(4,450,60)$. Fixed-delay networks train with $T_2=100$ and an adaptive temporal gate. Fixed-gate networks train with $T_2$ sampled uniformly from 50 to 150 steps and a shared Gaussian gate centered on the final retrieval step, with a learned width.

For the non-adaptive RNN condition, we reuse the six $H=405$ networks from the training experiments in \appref{app:s1}, with initialization seeds 101--104, 107, and 108. These are the largest RNNs in that size sweep; they use $N=3$ and $T_3=60$. The five extra units implement the parameter-budget matching to the $H=400$ \arnn{}s described in \appref{app:readout}. Here we analyze each run's best-validation checkpoint from the 60-epoch training history, whereas the delay-generalization curves in \appref{app:s1} use the final epoch-60 checkpoints. Reusing these networks tests whether the phase-coding outcome of removing adaptation recurs across training seeds.

Table~\ref{tab:training} summarizes the general task and optimization settings. Checkpoints are selected by validation accuracy, with accuracy-based early stopping for the new fixed-delay and fixed-gate runs. Fixed-gate seed 21003 was continued from epoch 120 to the training limit of epoch 220; its best-validation checkpoint at epoch 217 reaches $93.95\%$ validation accuracy and is used here. All six RNNs and all four runs of each new ablation condition enter the comparison irrespective of phase scores or task performance.

The primary comparison uses $T_2=100$ for every model (Table~\ref{tab:ablation_multiseed}). This is the training delay of the fixed-delay networks and lies within the training range of all variable-delay networks. Robust $R_\phi$ retains the definition in \appref{app:conditions}, evaluated at this shared delay. For within-trial phase structure, each model's frequency is calibrated independently rather than fixed by $T_3$. Calibration uses a dedicated batch of trials from the official MNIST test-image pool, with sampling and noise streams separate from the reporting trials. All models use the same 100-step pre-cue interval and frequency grid $f=k/800$, $k=8,\ldots,399$. For each candidate frequency, each present-digit logit is fitted as
\begin{equation}
z_d(t)=a_d+b_dt+u_d\sin(2\pi ft)+v_d\cos(2\pi ft)+\epsilon_d(t).
\end{equation}
The incremental fit score is $R^2_{\mathrm{osc}}=1-\mathrm{RSS}_{\mathrm{full}}/\mathrm{RSS}_{\mathrm{linear}}$, relative to an intercept-and-linear-trend fit. Each nonconstant present-item trace contributes equally to the calibration mean. The frequency maximizing this mean is fixed for subsequent reporting; neither phase separation nor retrieval accuracy enters frequency selection. The resulting period is $\widehat P=1/\widehat f$. The common grid resolves periods up to 100 steps; longer periods remain unresolved by this calibration interval.

On reporting trials, phases are obtained from $\operatorname{atan2}(v_d,u_d)$ at the calibrated frequency. The circular minimum-gap definition in \appref{app:robust-opc} gives $S_{\mathrm{free}}$, summarized by its median over trials whose present-digit fits all satisfy $R^2_{\mathrm{osc}}\geq0.5$. Their fraction is $C_{\mathrm{free}}$. Zero residual energy after the linear fit yields an undefined phase. The explicit trend term prevents a purely linear drift from qualifying as an oscillation. Accuracy and $R_\phi$ include all trials without phase-fit or correctness filtering. This analysis uses the same period search, window, and inclusion criterion for every model, including the main \arnn{} reference.

\begin{table}[!t]
\centering
\footnotesize
\setlength{\tabcolsep}{4pt}
\caption{Phase structure at the shared in-distribution delay $T_2=100$. Each model has a period $\widehat P$ selected on independent calibration trials, without using $T_3$ or phase separation. $S_{\mathrm{free}}$ and fit coverage $C_{\mathrm{free}}$ use held-out delay activity; $R_\phi$ retains the main-text definition. Accuracy uses native retrieval. Mean $\pm$ sample SD summarizes runs, with one conditional separation median per run. A star marks the upper search boundary: the period remains unresolved. Periods are in simulation steps. NA denotes no qualifying trials.}
\label{tab:ablation_multiseed}
\begin{tabular}{lcrrrrr}
\toprule
Seed & $N/H/T_3$ & $\widehat P$ & $R_\phi$ & $S_{\mathrm{free}}$ & $C_{\mathrm{free}}$ (\%) & Acc. (\%) \\
\midrule
\multicolumn{7}{l}{\textit{Main \arnn{} reference}} \\
main & 4/500/55 & 50.0 & 0.967 & 0.755 & 99.41 & 97.81 \\
\midrule
\multicolumn{7}{l}{\textit{Fixed delay}} \\
21001 & 3/450/65 & 57.1 & 0.395 & 0.618 & 49.76 & 98.57 \\
21002 & 3/500/65 & 50.0 & 0.358 & 0.284 & 18.50 & 98.39 \\
21003 & 4/550/65 & 61.5 & 0.882 & 0.542 & 60.03 & 98.06 \\
21004 & 4/450/60 & 53.3 & 0.645 & 0.500 & 54.13 & 98.00 \\
Mean $\pm$ SD & -- & -- & $0.570\pm0.244$ & $0.486\pm0.143$ & $45.60\pm18.55$ & $98.25\pm0.27$ \\
\midrule
\multicolumn{7}{l}{\textit{Fixed temporal gate}} \\
21001 & 3/450/65 & 36.4 & 0.180 & 0.258 & 25.19 & 97.61 \\
21002 & 3/500/65 & 36.4 & 0.136 & 0.182 & 9.73 & 97.17 \\
21003 & 4/550/65 & 32.0 & 0.151 & 0.222 & 15.13 & 94.36 \\
21004 & 4/450/60 & 32.0 & 0.185 & 0.188 & 6.61 & 95.23 \\
Mean $\pm$ SD & -- & -- & $0.163\pm0.024$ & $0.213\pm0.035$ & $14.16\pm8.15$ & $96.09\pm1.55$ \\
\midrule
\multicolumn{7}{l}{\textit{RNN without adaptation (\appref{app:s1})}} \\
101 & 3/405/60 & 88.9 & 0.295 & 0.291 & 89.37 & 97.76 \\
102 & 3/405/60 & 100.0$^{*}$ & 0.404 & 0.272 & 92.17 & 97.89 \\
103 & 3/405/60 & 100.0$^{*}$ & 0.504 & 0.306 & 92.17 & 98.07 \\
104 & 3/405/60 & 100.0$^{*}$ & 0.405 & 0.294 & 89.71 & 97.94 \\
107 & 3/405/60 & 100.0$^{*}$ & 0.377 & 0.320 & 94.70 & 97.37 \\
108 & 3/405/60 & 100.0$^{*}$ & 0.475 & 0.349 & 89.87 & 97.79 \\
Mean $\pm$ SD & -- & -- & $0.410\pm0.074$ & $0.305\pm0.027$ & $91.33\pm2.07$ & $97.80\pm0.24$ \\
\bottomrule
\end{tabular}
\end{table}

Delay generalization is evaluated separately at $T_2=220$, the extended maintenance duration in \figref{fig:conditions}a(iii), using another forward pass of the same reporting trials (Table~\ref{tab:ablation_persistence}). An independent calibration batch at this delay selects $\widehat P_{220}$ using the same mean incremental-fit criterion, with $f=k/(8\times220)$, $k=8,\ldots,879$. This longer interval admits longer periods without constraining them to the native-delay estimate. The selected frequency is held fixed across two adjacent 110-step maintenance windows, both preceding the cue. $C_{\mathrm{all}}$ is the fraction of trials for which every presented item passes the fit criterion in both windows. Among these trials, $S_{\min}$ summarizes the minimum separation across the two windows. Relative-phase consistency is computed for each trial as
\begin{equation}
C_{\Delta,\ell}=\binom{N}{2}^{-1}\sum_{i<j}\left|\frac{1}{2}\sum_{w=1}^{2}
\exp\!\left[\mathrm{i}(\phi_{\ell iw}-\phi_{\ell jw})\right]\right|.
\end{equation}
We report its median across qualifying trials. It equals one when pairwise phase differences remain unchanged across windows, including the case of coincident phases. Amplitude retention is the within-trial median, across presented items, of second-window amplitude divided by first-window amplitude, then the median across qualifying trials; amplitude is $\sqrt{u_d^2+v_d^2}$. These conditional summaries are interpreted alongside $C_{\mathrm{all}}$ and measure persistence at the calibrated frequency.

\begin{table}[!t]
\centering
\footnotesize
\setlength{\tabcolsep}{4pt}
\caption{Persistence at the extended delay of Fig.~\ref{fig:conditions}a(iii), $T_2=220$. The period $\widehat P_{220}$ is selected on independent calibration trials at this delay. A star marks the upper search boundary, not a resolved oscillation period. Both adjacent 110-step pre-cue windows use this frequency. $C_{\mathrm{all}}$ is the percentage of trials passing every present-item fit in both windows. Conditional on this criterion, $S_{\min}$ is median minimum separation across windows, $C_{\Delta}$ is median relative-phase consistency, and $A_2/A_1$ is median amplitude retention. Parentheses count runs with defined conditional metrics. Periods are in simulation steps. This is a separate delay-generalization test.}
\label{tab:ablation_persistence}
\begin{tabular}{lrrrrr}
\toprule
Seed & $\widehat P_{220}$ & $C_{\mathrm{all}}$ (\%) & $S_{\min}$ & $C_{\Delta}$ & $A_2/A_1$ \\
\midrule
\multicolumn{6}{l}{\textit{Main \arnn{} reference}} \\
main & 51.8 & 99.60 & 0.763 & 0.994 & 1.38 \\
\midrule
\multicolumn{6}{l}{\textit{Fixed delay}} \\
21001 & 58.7 & 46.69 & 0.565 & 0.906 & 1.56 \\
21002 & 58.7 & 7.06 & 0.407 & 0.911 & 1.65 \\
21003 & 60.7 & 55.91 & 0.516 & 0.972 & 2.10 \\
21004 & 62.9 & 11.40 & 0.413 & 0.961 & 1.50 \\
Mean $\pm$ SD & -- & $30.26\pm24.64$ & $0.475\pm0.078$ & $0.937\pm0.034$ & $1.70\pm0.27$ \\
\midrule
\multicolumn{6}{l}{\textit{Fixed temporal gate}} \\
21001 & 38.3 & 21.11 & 0.129 & 0.959 & 1.09 \\
21002 & 73.3 & 1.69 & 0.045 & 0.953 & 0.65 \\
21003 & 60.7 & 0.03 & 0.165 & 0.976 & 1.08 \\
21004 & 35.2 & 0.66 & 0.089 & 0.895 & 1.19 \\
Mean $\pm$ SD & -- & $5.87\pm10.18$ & $0.107\pm0.052$ & $0.946\pm0.036$ & $1.00\pm0.24$ \\
\midrule
\multicolumn{6}{l}{\textit{RNN without adaptation (\appref{app:s1})}} \\
101 & 220.0$^{*}$ & 66.73 & 0.129 & 0.626 & 0.09 \\
102 & 220.0$^{*}$ & 87.17 & 0.186 & 0.711 & 0.38 \\
103 & 220.0$^{*}$ & 78.83 & 0.140 & 0.609 & 0.07 \\
104 & 220.0$^{*}$ & 78.10 & 0.158 & 0.656 & 0.11 \\
107 & 220.0$^{*}$ & 87.51 & 0.165 & 0.680 & 0.33 \\
108 & 220.0$^{*}$ & 74.90 & 0.164 & 0.633 & 0.09 \\
Mean $\pm$ SD & -- & $78.87\pm7.84$ & $0.157\pm0.020$ & $0.652\pm0.038$ & $0.18\pm0.14$ \\
\bottomrule
\end{tabular}
\end{table}

At the trained delay, the fixed-delay networks reach $98.25\%\pm0.27\%$ accuracy and exhibit phase separation with mean within-run median $S_{\mathrm{free}}=0.486\pm0.143$. Thus, fixed-delay training can produce phase structure at the trained timing. At $T_2=220$, their calibrated periods are 58.7--62.9 steps, while the main \arnn{} has $\widehat P_{220}=51.8$ steps. Across the four fixed-delay runs, $C_{\mathrm{all}}$ averages $30.26\%$, mean conditional $S_{\min}$ is $0.475$, and mean $C_{\Delta}$ is $0.937$, compared with $99.60\%$, $0.763$, and $0.994$ for the main \arnn{}. Persistence varies across seeds: fixed-delay seed 21003 retains qualifying fits in $55.91\%$ of trials, with conditional $S_{\min}=0.516$ and $C_{\Delta}=0.972$. Fixed-delay training therefore yields less consistently separated phase structure across the extended maintenance interval, while a subset of trials retains stable phase relationships. These measurements distinguish sustained within-trial separation from the presence of oscillatory activity alone.

The six non-adaptive RNNs achieve $97.80\%\pm0.24\%$ accuracy at $T_2=100$, with $C_{\mathrm{free}}=91.33\%\pm2.07\%$ and $S_{\mathrm{free}}=0.305\pm0.027$. Five select the 100-step upper period boundary. Extending the calibration interval to 220 steps moves all six estimates to the new upper boundary, rather than revealing an interior oscillatory timescale comparable to the main \arnn{}. These values are marked with a star in the tables: they remain unresolved periods and are consistent with slow modulation over the observation interval rather than clearly repeated oscillatory cycles. Under the extended delay, mean conditional $S_{\min}$ is $0.157$, $C_{\Delta}$ is $0.652$, and amplitude retention is $0.18$. For boundary estimates, these phase quantities characterize finite-window fits rather than demonstrated periodic activity. Across all four fixed-gate runs, native-delay calibrated periods are 32.0--36.4 steps and $C_{\mathrm{free}}$ is $6.61\%$--$25.19\%$; their conditional separation medians are $0.182$--$0.258$. Mean test accuracy is $96.09\%\pm1.55\%$, with mean $S_{\mathrm{free}}=0.213\pm0.035$ and $C_{\mathrm{free}}=14.16\%\pm8.15\%$. For the continued seed 21003, test accuracy is $94.36\%$ and $C_{\mathrm{free}}=15.13\%$ at $T_2=100$. At $T_2=220$, its calibrated period is $60.7$ steps, but just 2 of 7,000 trials pass every present-item fit in both windows ($C_{\mathrm{all}}=0.03\%$). Its conditional $S_{\min}=0.165$, $C_{\Delta}=0.976$, and amplitude retention $1.08$ describe these two trials. Across the four fixed-gate runs, mean $C_{\mathrm{all}}$ is $5.87\%$ and mean conditional $S_{\min}$ is $0.107$.

Group summaries report means and sample standard deviations across independently trained networks, using one conditional median per network. Runs without qualifying trials retain zero coverage and undefined conditional metrics. The main \arnn{} row is a single-checkpoint reference. Network sizes, task configurations, and training budgets differ across cohorts.

\section{The hand-designed content--time circuit}
\label{app:circuit}
\subsection{Circuit dynamics and simulation settings}
This section specifies the matrices, inputs, and numerical updates in Eqs.~\eqref{eq:circuit-content}--\eqref{eq:circuit-ring-rate}, using the notation of \secref{sec:constructive}. The simulations use $N_C=10$ content units, one per digit identity, and $N_T=60$ time units arranged in a ring. Content activity is $r_{C,i}=\max(\tanh U_i^C,0)$; the ring input to each content unit is gated by its own activity, $g_i=G(r_{C,i})$; time-ring activity $r_{T,j}$ follows Eq.~\eqref{eq:circuit-ring-rate}. Both $U^P$ and $V^P$ are signed states for population $P\in\{C,T\}$. All matrix products use column activity vectors, so a matrix's first index denotes the receiving unit.

\paragraph{Recurrent connectivity and ring dynamics.}
The content matrix is $W_C=w_{\mathrm{self}}I+J^C$, with $w_{\mathrm{self}}=0.1$. Its random component $J^C$ is sampled as specified below. For the time ring, $\varphi_j=2\pi j/N_T$. The rate normalization uses the effective density $\rho=2\pi/N_T$ (the product of the neural density and the angular spacing) and $\kappa=0.134$, with the square applied to the signed state $U_j^T$; $\rho$ is the per-unit numerical integration weight in the ring sums. The symmetric kernel in $W_T$, including this weight, is
\begin{equation}
(W_T)_{j\ell}=\frac{J_0\rho}{\sqrt{2\pi}\sigma_{\mathrm w}}
\sum_{p\in\mathbb Z}\exp\!\left[-\frac{(\varphi_j-\varphi_\ell+2\pi p)^2}{2\sigma_{\mathrm w}^2}\right],
\qquad \sigma_{\mathrm w}=0.35,\quad J_0=1,
\end{equation}
with kernel width $\sigma_{\mathrm w}$ in radians and overall coupling gain $J_0$, implemented by a finite periodic-image sum. Thus $(W_T\mathbf r_T)_j=\sum_\ell(W_T)_{j\ell}r_{T,\ell}$. The component form of Eq.~\eqref{eq:circuit-time} and its adaptation equation is
\begin{equation}
\begin{aligned}
\tau_U^T\dot U_j^T&=-U_j^T+\sum_\ell(W_T)_{j\ell}r_{T,\ell}+I_{C\to T,j}+I_j^T-V_j^T,\\
\tau_V^T\dot V_j^T&=-V_j^T+m^T U_j^T,
\end{aligned}
\end{equation}
where $I_{C\to T,j}$ denotes the content feedback term expanded below. The time constants are $\tau_U^T=0.1$ and $\tau_V^T=6$, and the adaptation strength is $m^T=0.08$. This is the Gaussian adaptation model associated with traveling activity in continuous-attractor networks \citep{mi2014}, combined here with content coupling. No asymmetric sine kernel or subtractive population-inhibition term is added. Parameter values are reported rounded; the exact values are given in the simulation code (see the Reproducibility statement).

\paragraph{Inputs to content units.}
The complete drive inside the $\tanh$ nonlinearity for content unit $i$ in Eq.~\eqref{eq:circuit-content} is
\begin{equation}
\begin{aligned}
I_i^{\mathrm{tot},C}(t)
&=(W_C\mathbf r_C)_i+\lambda g_i(t)(W_{T2C}\mathbf r_T)_i+I_i^C(t)\\
&=w_{\mathrm{self}}r_{C,i}(t)+\sum_\ell J^C_{i\ell}r_{C,\ell}(t)+\lambda g_i(t)S(t)+I_i^C(t),
\end{aligned}
\label{eq:content-input}
\end{equation}
Here $g_i=G(r_{C,i})$, and the shared signal satisfies $(W_{T2C}\mathbf r_T)_i=S$ for every content unit. The external input in the main-text equation is $I_i^C=I_i^{\mathrm{digit}}+\zeta_i^C$. It contains the encoding pulse and noise; recurrent and interpopulation currents are the separate terms shown in Eq.~\eqref{eq:content-input}. For a sequence $(d_1,\ldots,d_N)$ presented for $D$ steps per item,
\begin{equation}
I_i^{\mathrm{digit}}(t)=A\sum_{k=1}^N
\mathbf 1_{\{i=d_k\}}\mathbf 1_{\{(k-1)D\leq t<kD\}}.
\end{equation}
The simulations use $A=1.2$, $D=10$, and $N=4$. This pulse is zero throughout maintenance. The noise $\zeta_i^C$ has independent Gaussian samples with standard deviation $0.02$ at each outer step, with an additional independent component of standard deviation $0.004$ during maintenance.

\paragraph{Projection matrices and activity-dependent coupling.}
Time-ring activity is pooled into a shared signal $S$ through four local receptive fields:
\begin{equation}
\begin{aligned}
B_{kj}&=\exp[20(\cos(\varphi_j-\xi_k)-1)],\\
q_k(t)&=\frac{\sum_jB_{kj}r_{T,j}(t)}{\sum_jB_{kj}+10^{-12}},
\qquad S(t)=\sum_kq_k(t),
\end{aligned}
\end{equation}
where the field centers are $\xi_k=2\pi k/4$ for $k=0,1,2,3$. Each $q_k$ averages activity near one ring location; their sum supplies the same signal to all content units. These are readouts, not extra simulated neural populations. The corresponding main-text projection matrices are
\begin{equation}
(W_{T2C})_{ij}=\sum_{k=0}^{3}\frac{B_{kj}}{\sum_\ell B_{k\ell}+10^{-12}},
\qquad (W_{C2T})_{ji}=1.
\label{eq:circuit-projections}
\end{equation}
Thus $W_{T2C}\in\R^{N_C\times N_T}$ has identical rows and $W_{C2T}\in\R^{N_T\times N_C}$ broadcasts the content sum to every time unit. With $s(x)=(1+\tanh(10x))/2$, the gate in the main-text equations is
\begin{equation}
g_i=G(r_{C,i})=\frac{\max\{s(r_{C,i}-0.25)-s(-0.25),0\}}{1-s(-0.25)}.
\end{equation}
Thus $G(0)=0$, and the time-ring input increases smoothly as the content unit becomes active. The same gate function is applied to every content unit, using its current activity. Encoding pulses establish different activation times, and the resulting activity-dependent drive supports their subsequent phase-separated reactivation.

Substituting these matrices and gates into Eqs.~\eqref{eq:circuit-content} and \eqref{eq:circuit-time} gives the reciprocal coupling currents
\begin{equation}
 I_{T\to C,i}=\lambda g_iS,
 \qquad I_{C\to T,j}=\eta\tanh\!\left(\sum_i g_i r_{C,i}\right).
 \label{eq:coupling}
\end{equation}
Here $\lambda\simeq1.07$ and $\eta=0.5$ set the coupling strengths; parameter robustness around this operating point is evaluated in \appref{app:circuit-parameters}. Content feedback pools the gated content activities $g_i r_{C,i}$ and is uniform over ring positions. The external time-ring input $I_j^T$ combines the moving encoding input with noise, as specified below. Divisive normalization acts on $\mathbf r_T$ before the projection $W_{T2C}\mathbf r_T$. The time-to-content current enters the content unit's $\tanh$, whereas the saturated content-to-time current is added directly to the time-unit dynamics.

\paragraph{Numerical updates and simulation protocol.}
For the continuous-time content dynamics in Eq.~\eqref{eq:circuit-content}, the simulation uses $\tau_U^C=10$, $\tau_V^C=1/0.014$, and $m^C=20$, with outer step size $\Delta t=1$. A sequential first-order scheme updates $U_i^C$ first and then evaluates the adaptation drive at the updated state $U_{i,t+1}^C$. In the notation of the main text, these updates are
\begin{align}
U_{i,t+1}^C&=0.9U_{i,t}^C+0.1\tanh\!\left(I_{i,t}^{\mathrm{tot},C}\right)-0.1V_{i,t}^C,\\
V_{i,t+1}^C&=0.986V_{i,t}^C+0.28U_{i,t+1}^C.
\end{align}
Here $I_{i,t}^{\mathrm{tot},C}$ is evaluated from pre-update activities using Eq.~\eqref{eq:content-input}.
The cross-talk matrix $J^C$ is drawn with entry standard deviation $0.02$, has its diagonal set to zero, and is then centered by subtracting its matrix-wide mean. Ring integration uses 80 simultaneous Euler substeps of size $1/80$ per outer step. Each substep updates $U^T,V^T$ using their previous values, while external input and content feedback stay fixed across the outer step. Content-dependent coupling gates and currents are computed from pre-update states; displayed activities are post-update values.

During the first 40 steps, the four content units receive sequential ten-step currents of amplitude $1.2$. Ring entrainment uses the external input
\begin{equation}
I_j^T(t)=3\exp\!\left[-\frac{\operatorname{wrap}(\varphi_j-\omega t)^2}{4\sigma_{\mathrm w}^2}\right]
\mathbf 1_{\{0\leq t<40\}}+\zeta_j^T(t),
\label{eq:circuit-ring-input}
\end{equation}
where $\operatorname{wrap}$ maps angular differences to $[-\pi,\pi)$, $\omega=0.153$ radians per outer step, and the initial input center is zero. Both deterministic input sequences end at maintenance onset. The noise $\zeta_j^T$ is sampled independently at each outer step with standard deviation $0.01$ and held constant during its ring substeps. Initial $U_j^T$ and $U_i^C$ are zero-mean Gaussian with standard deviation $0.02$; both adaptation states start at zero. The same random realization is reset for each of the 5,040 distinct ordered sequences, fixing the circuit, initial conditions, and noise realization.

The main-text population simulation ends at step 190, after 40 encoding and 150 maintenance steps. The circuit parameters are fixed across all sequences. The time-ring heatmap averages rates across all sequences over steps $[110,190)$. A period-40 harmonic fit to each mean time-unit trace determines row ordering; all 60 time units remain. The reference period supplies a phase coordinate without forcing the simulated activity to have exactly that period. Its narrow pulses need not be sinusoidal, and its nonnegative color scale differs from the z-scored hidden-neuron heatmap in \figref{fig:cells}d.

\subsection{Oscillatory phase coding in the circuit}
\label{app:circuit-opc}
\begin{figure}[!t]
\centering
\includegraphics[width=\linewidth]{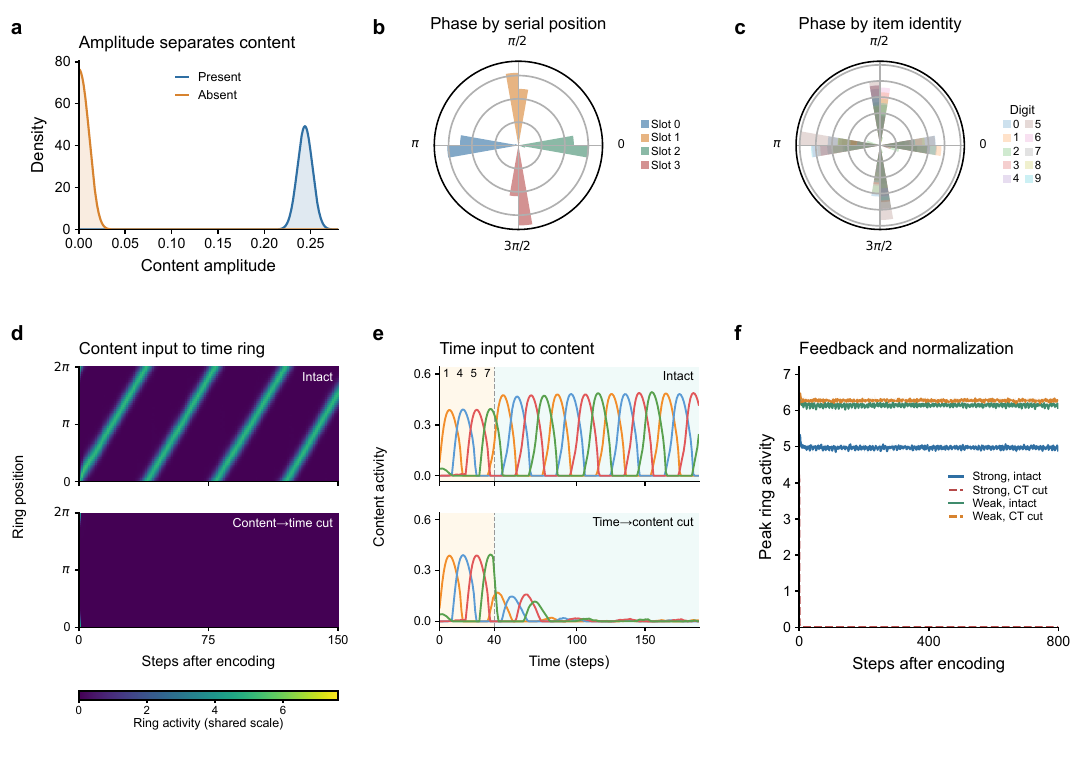}
\caption{\textbf{Reciprocal coupling sustains content and time-ring activity.} (a) Oscillation-amplitude distributions for present and absent content units. (b,c) Phases separate by serial position (b) but overlap by item identity (c). (d) Time-ring activity with intact (top) or removed (bottom) content-to-time feedback under strong divisive normalization, shown on a common color scale. (e) Activity of the presented content units in \figref{fig:circuit}b with intact (top) or removed (bottom) time-to-content input, shown on a common activity scale. Connections are removed at maintenance onset. (f) Peak ring activity with intact or removed content-to-time feedback under strong and weak normalization. Lines and shading show means and SDs across three simulation runs.}
\label{fig:circuit-support}
\end{figure}
Across all 5,040 ordered four-item sequences, present content units have mean harmonic amplitude $0.24382$, compared with $0.00578$ for absent units (\figref{fig:circuit-support}a). Demeaned content activity is fitted with sine and cosine terms at a reference period of 40 steps. Amplitudes use steps $[150,190)$ and are averaged within each sequence over present or absent items. Density curves use Gaussian kernels with a common absolute bandwidth of $0.008$, reflected at zero.

Phases use the longer maintenance window $[110,190)$ and the same harmonic convention. All presented-item fits meet $R^2\geq0.5$. One global rotation places the mean phase of the first position at $\pi$; no item-specific alignment is applied. The same phase estimates are grouped by serial position or by item identity in \figref{fig:circuit-support}b,c. The 36-bin polar histograms are normalized densities. Concentration by position ranges from $0.996$ to $0.997$, whereas concentration by item identity ranges from $0.004$ to $0.023$. The distinct sequences probe a single fixed circuit with matched initialization and noise, rather than independently trained networks.

\subsection{Reciprocal coupling and bump maintenance}
\label{app:circuit-feedback}
Matched feedback interventions show how normalization controls the ring's dependence on content. For each of three stochastic repetitions, the intact circuit encodes the same four-item sequence. At maintenance onset, copies of its content state, ring state, and both adaptation states branch into four conditions: intact or removed content-to-time feedback, each at strong or weak divisive normalization. All branches share encoding, cross-talk, and subsequent external-input noise. Only maintenance parameters change. Strong normalization uses the principal $\kappa=0.134$; weak normalization uses $\kappa=0.121$. Cutting feedback sets its coupling gain to zero.

The feedback simulations contain 40 encoding and 4,000 maintenance steps. Panel d shows the first 150 maintenance steps of one matched repetition, with intact feedback above removed feedback, in the original spatial ring order and on a common activity scale. Panel f shows peak activity across ring units over the first 800 maintenance steps; lines and shading give the mean and sample SD across the three repetitions. In the final 150 steps of this displayed window, the strong intact circuit has mean peak activity $4.972$, whereas feedback removal reduces it to approximately $6.8\times10^{-4}$. Under weaker normalization, the corresponding intact and cut values are $6.148$ and $6.272$. Thus the same feedback removal has different effects depending on normalization strength, supporting reciprocal maintenance in the principal strongly normalized circuit (\figref{fig:circuit-support}d,f).

The complementary intervention in \figref{fig:circuit-support}e removes time-to-content input by setting $\lambda=0$ at maintenance onset. It uses the same 40-step encoding and 150-step maintenance simulation as \figref{fig:circuit}b, with encoding states, content connections, and subsequent noise matched between the intact and cut conditions. The presented content units retain oscillatory activity with intact coupling, whereas their oscillation amplitudes decay after the input is removed. Over steps $[150,190)$, their mean harmonic amplitude decreases from $0.24490$ to $0.00260$, close to the absent-unit amplitude of $0.00254$ in the cut condition. Together, the two interventions show that reciprocal input sustains both populations in the principal circuit.

\subsection{Parameter robustness of reciprocal coupling}
\label{app:circuit-parameters}
We scan the two reciprocal coupling gains while keeping the gate function, recurrent content matrix, and all other circuit parameters fixed. Each gain takes 21 logarithmically spaced values from 0.1 to 10 times its reference value, including the reference pair $(\lambda_0,\eta_0)\simeq(1.07,0.5)$. We evaluate both normalization strengths used in \appref{app:circuit-feedback}, giving 882 grid points. Each point uses 20 distinct four-item sequences, balanced so that each digit occurs twice at each serial position, and three noise/initial-state realizations. The same sequences and realizations are matched across parameter settings; the content connection matrix remains that of the principal circuit. Each simulation contains 40 encoding and 150 maintenance steps, with the scanned parameters applied throughout.

\begin{figure}[!t]
\centering
\includegraphics[width=\linewidth]{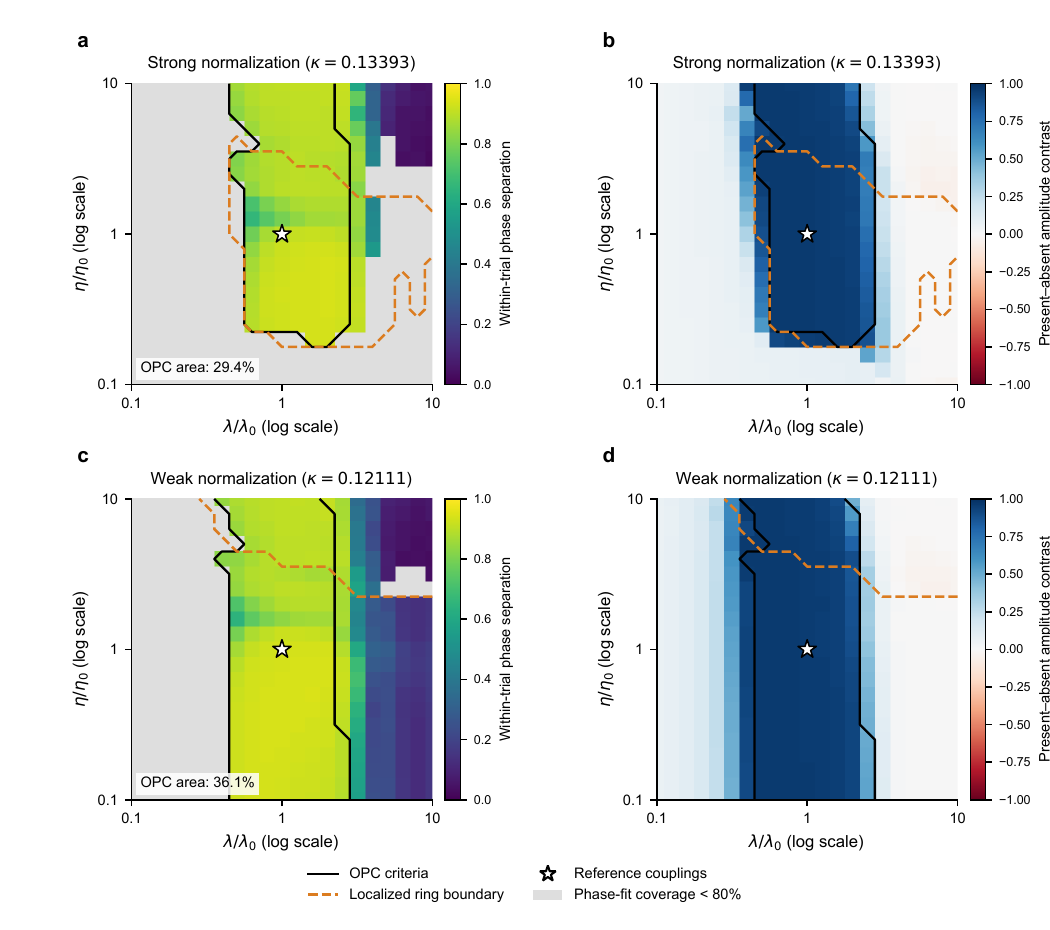}
\caption{\textbf{Phase coding occupies a connected parameter region around the reference couplings.} Within-trial phase separation (a,c) and present--absent amplitude contrast (b,d) over the reciprocal coupling gains, under strong (a,b) or weak (c,d) normalization. Black contours delimit the operational OPC region; dashed orange contours mark localized ring activity. Stars mark the reference coupling pair in each slice. Gray phase-map cells have less than $80\%$ complete-phase fit coverage. Area percentages refer to the scanned logarithmic coordinates.}
\label{fig:circuit-parameters}
\end{figure}

Measurements use steps $[110,190)$. At each grid point, a shared frequency is estimated from the mean spectrum of the presented content units across all 60 simulations. Traces are linearly detrended, normalized by their residual energy, Hann-windowed, and transformed on an eightfold zero-padded grid; the search admits periods from just above two to 80 steps, with log-parabolic refinement of an interior peak. Phase and amplitude are then estimated by an intercept, linear trend, and sine/cosine fit at this frequency. A trial has complete-phase coverage when every presented item has incremental $R^2\geq0.5$ relative to the linear-only fit. Separation $S$ uses the normalized minimum circular gap defined in \appref{app:robust-opc}. Across qualifying trials, pairwise phase consistency is the mean over item pairs of $|\langle e^{\mathrm i(\phi_i-\phi_j)}\rangle|$. Amplitude contrast is $D_A=(\overline A_{\mathrm{present}}-\overline A_{\mathrm{absent}})/(\overline A_{\mathrm{present}}+\overline A_{\mathrm{absent}})$, using amplitudes from all simulations.

The operational OPC region requires complete-phase coverage of at least $0.8$, median $S\geq0.5$, pairwise consistency of at least $0.8$, $D_A\geq0.5$, and mean present-item amplitude of at least $0.05$. The amplitude-contrast criterion corresponds to a present/absent ratio of at least three; the absolute-amplitude floor excludes negligible signals. The region occupies $29.4\%$ of the scanned log-coordinate area under strong normalization and $36.1\%$ under weak normalization (\figref{fig:circuit-parameters}). Areas are trapezoidal grid estimates over $[-1,1]^2$ in $(\log_{10}(\lambda/\lambda_0),\log_{10}(\eta/\eta_0))$, not fractions of a linear-coordinate domain. All qualifying grid points belong to the component containing the reference pair. Loosening the five criteria to $(0.7,0.4,0.7,0.4,0.025)$ gives areas of $30.4\%$ and $38.3\%$; tightening them to $(0.9,0.6,0.9,0.6,0.075)$ gives $28.4\%$ and $34.6\%$.

Around the strongly normalized reference, all 25 sampled combinations in the centered $5\times5$ block, spanning 0.631--1.585 times each coupling gain, satisfy the OPC criteria. Under weak normalization, the corresponding block contains 49 combinations spanning 0.501--1.995 times each gain. A separate simulation with the rounded values $\lambda=1.07$, $\eta=0.5$, and $\kappa=0.134$ also satisfies the stricter criteria, with median separation $0.900$ and amplitude contrast $0.954$.

Stronger normalization raises the feedback requirement for maintaining time-ring activity. At the reference time-to-content coupling $\lambda=\lambda_0$, the smallest sampled $\eta$ satisfying the OPC criteria is $0.251\eta_0$ under strong normalization, whereas weak normalization supports OPC down to the scan limit of $0.1\eta_0$. At $\eta=0.1\eta_0$, the strongly normalized ring decays to mean peak activity $0.00069$ and complete-phase coverage is zero; under weak normalization, the ring retains localized activity with mean peak $6.29$ and complete-phase coverage is $100\%$. The upward shift of the lower feedback boundary supports a maintenance role for content-to-time input in the strongly normalized circuit. Separately, excessive time-to-content coupling ($\lambda=10\lambda_0$ at $\eta=\eta_0$) recruits absent content units and eliminates amplitude selectivity in both normalization settings.

Increasing $\eta$ supplies a larger spatially uniform drive to the ring. The dashed contours mark spatial localization $|\sum_j r_{T,j} e^{\mathrm i\varphi_j}|/\sum_j r_{T,j}\geq0.2$, together with mean peak activity above $0.05$. At $\lambda=\lambda_0$, increasing $\eta$ from $0.5$ to $5$ reduces median ring localization from $0.912$ to $0.00165$ in the strong-normalization slice, while content phase separation remains $0.904$. Thus a change in ring dynamics can occur within the OPC-supporting region; the output-coding contour and the localization boundary describe different properties. These maps assess parameter robustness over the same finite maintenance interval as the main circuit demonstration.

\section{Learning and delay generalization across model sizes}
\label{app:s1}
\begin{figure}[htbp]
\centering
\includegraphics[width=\linewidth]{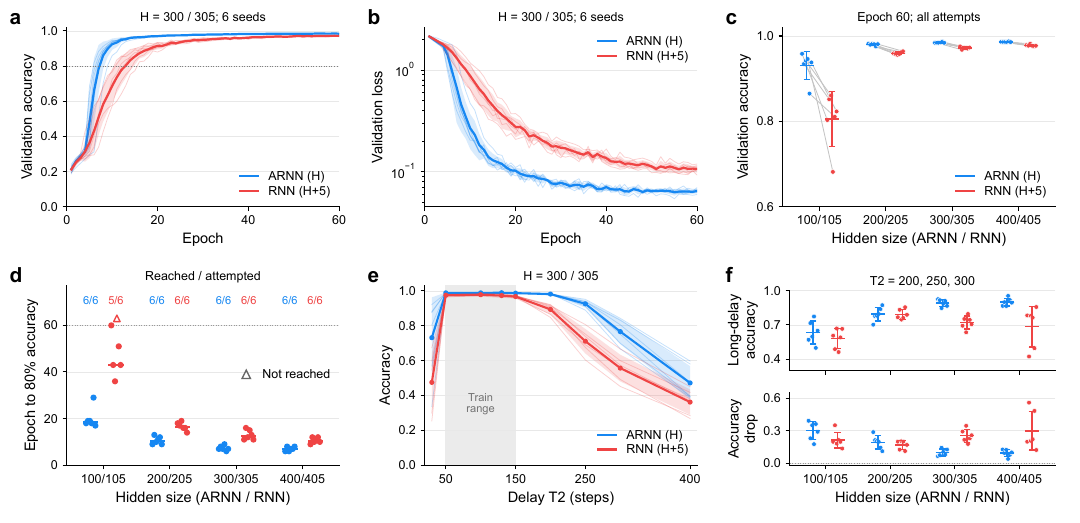}
\caption[Learning and delay generalization of adaptive and non-adaptive RNNs]{\input{sections/s1_caption}}
\label{fig:s1}
\end{figure}

We compare adaptive and non-adaptive recurrent networks on the same three-item task across four size settings (\figref{fig:s1}). Each \arnn{} has $H\in\{100,200,300,400\}$ hidden units; its RNN comparison has $H+5$ units without adaptation. The additional RNN units implement the approximate parameter-budget matching described in \appref{app:readout}. Six networks with different random initialization seeds were trained per architecture and size, for 48 runs in total. Every run completed 60 epochs without early stopping, and all runs enter the plots.

Training uses 75,000 trials per epoch and 15,000 validation trials, batch size 64, and fixed training and validation data sampling. Three digits are each presented for ten steps over two cycles ($T_1=60$); $T_3=60$ and training delays range from 50 to 150 steps. The learning-rate schedule uses five warmup epochs and 55 cosine-decay epochs. The initial base learning rate is $1.2\times10^{-4}$ and the adaptation-parameter learning rate is $2\times10^{-3}$. Panels a--d use the recorded validation histories; panel c uses epoch 60. Panels e,f evaluate each run's final checkpoint at epoch 60 at nine fixed delays, using 1,024 trials per delay. The reported accuracy values are rounded to three decimal places. In panels a,b,e, thin lines represent individual networks and thick lines and bands show their mean and sample SD. Panels a,b,e display the $H=300/305$ comparison; loss uses a logarithmic axis. Panel c pairs runs with matching initialization seeds across architectures, with mean and SD marks. Panel d reports the first epoch reaching 80\% accuracy, the median among networks that reach it, and open triangles for runs that do not reach it within 60 epochs. The lower part of panel f subtracts each model's long-delay mean from its accuracy at $T_2=130$. All error bars and bands describe variation across trained networks. For the long-delay summary, the three delays 200, 250, and 300 are averaged within each model before calculating means and sample SDs across seeds.

\arnn{}s reach 80\% validation accuracy in all 24 runs; the RNN reaches it in 23 of 24, with one $H+5=105$ run remaining below threshold after 60 epochs. \arnn{}s have higher mean epoch-60 validation accuracy at every tested size. At $H=300/305$, mean long-delay accuracy is 0.892 for \arnn{}s and 0.721 for RNNs. At $H=200/205$, the corresponding means are similar (0.795 and 0.792), so the long-delay advantage varies with network size. The tested sweep contains no $H=500$ group.

\section{Oscillatory activity and robustness across trained networks}
\label{app:robustness}
We first characterize the hidden neurons of the principal model, then examine whether its phase coding and cell populations recur across independently trained networks. Each trained network is the unit of replication.

\subsection{Adaptation and oscillatory activity of hidden neurons}
\label{app:oscillator-properties}
\begin{figure}[htbp]
\centering
\includegraphics[width=\linewidth]{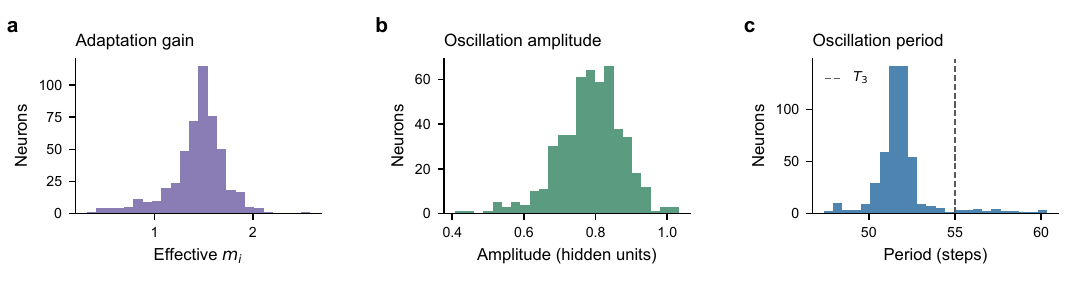}
\caption{\textbf{Adaptation strengths and oscillation properties of hidden neurons.} Distributions of learned adaptation strength (a), oscillation amplitude (b), and period (c) across all 500 hidden neurons. The dashed line marks the retrieval duration, $T_3=55$ steps.}
\label{fig:oscillator-properties}
\end{figure}
All 500 neurons are included. Effective gains are computed as $m_i=\operatorname{softplus}(\widetilde m_i)+10^{-6}$, and the dashed reference in \figref{fig:oscillator-properties}c marks $T_3=55$ steps. The effective adaptation gains have median $1.481$ and interquartile range $1.320$--$1.592$. A new evaluation of the same checkpoint fixes $T_2=135$ and analyzes its final $T_3=55$ pre-cue steps. For each trial and neuron, activity is demeaned, transformed with a 2,048-point real FFT, and normalized by its positive-frequency power. These spectra are averaged across trials separately for each neuron. We search the full positive-frequency grid and refine an interior peak by log-parabolic interpolation. No window taper or common-mode subtraction is applied to these spectra. The median estimated period is $51.70$ steps. Zero-padding refines the interpolation grid without improving the native frequency resolution of the 55-step observation window.

For amplitude, each trial is fitted with sine and cosine terms at that neuron's estimated frequency. Its median amplitude across trials yields one value for \figref{fig:oscillator-properties}b, in the original hidden-activity units. The population median is $0.791$, with interquartile range $0.737$--$0.844$; no per-neuron variance normalization is used.

\subsection{Robustness of oscillatory phase coding}
\label{app:robust-opc}
We analyze one selected checkpoint for each of twelve independently seeded task configurations, varying sequence length $N\in\{3,4,5\}$, retrieval duration $T_3\in\{60,65,70\}$, and hidden size $H\in\{450,500,550\}$ (Table~\ref{tab:multiseed-runs}). Within each four-seed block, sequence lengths comprise one three-item, two four-item, and one five-item task. Images from the official MNIST training set were split into 50,000 training and 10,000 validation images. Checkpoints are selected by validation accuracy, choosing the earlier epoch in a tie. Nine original runs reach validation accuracy $\geq0.95$. Continuing a10 from epoch 80 to epoch 170 yields a best checkpoint at epoch 165 with validation accuracy $97.86\%$. For a03 and a07, we retrain from the same initialization seeds for 120 epochs with a longer learning-rate schedule, then continue their best checkpoints with warmup and cosine annealing until accuracy plateaus. The selected checkpoints are a03 epoch 253 ($96.18\%$) and a07 epoch 244 ($95.19\%$). All twelve settings are now represented: three three-item, six four-item, and three five-item networks. The original runs share the same base optimization configuration: a peak learning rate of $2\times10^{-4}$ for ordinary parameters, $2\times10^{-3}$ for adaptation parameters, and a common warmup and cosine schedule. All three five-item runs initially remained below the task-accuracy criterion, whereas the three- and four-item runs reached it. This pattern suggests greater optimization demands at $N=5$. The subsequent training uses longer budgets and adjusted learning rates; the a03/a07 retraining doubles the ordinary-parameter peak to $4\times10^{-4}$ before their later gentle continuation. These outcomes suggest that extended training and learning-rate adjustment can help meet the greater optimization demands of the five-item task. Table~\ref{tab:multiseed-runs} reports the resulting checkpoints. Inclusion uses task performance without filtering on phase scores or independent-test results.

\begin{figure}[!t]
\centering
\includegraphics[width=\linewidth]{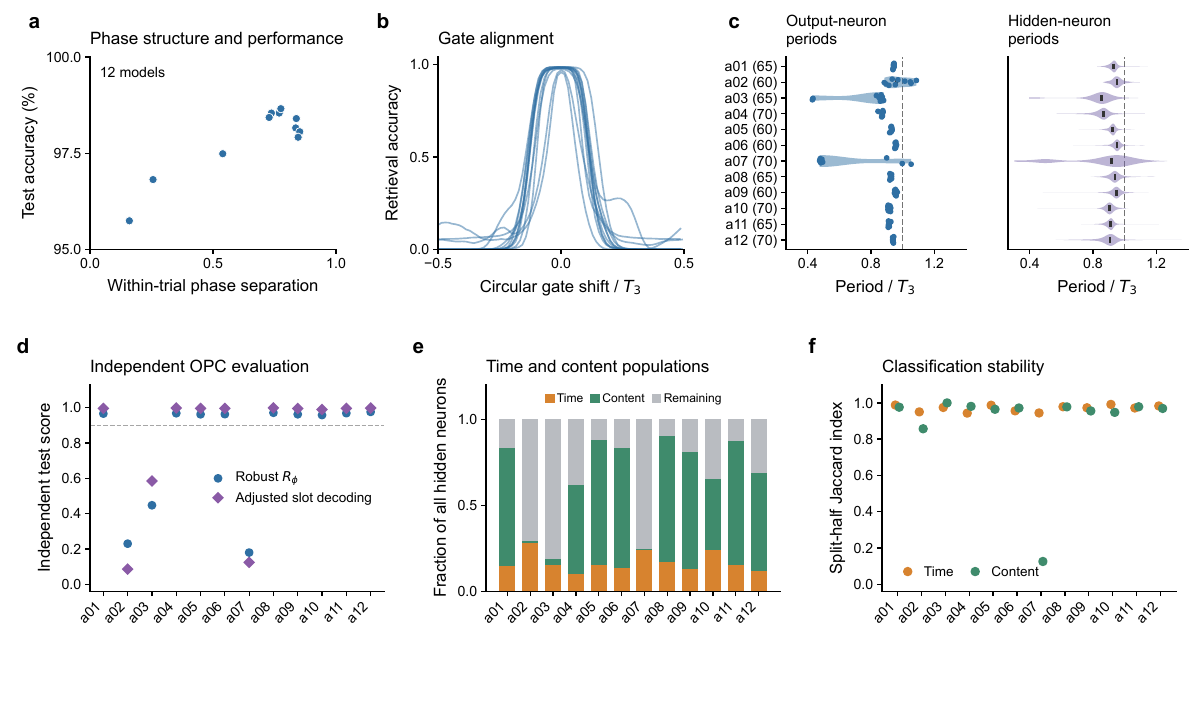}
\caption{\textbf{Phase coding and population organization across independently trained networks.} (a) Within-trial phase separation and task performance. (b) Retrieval depends on gate alignment. (c) Output-neuron and hidden-neuron periods across models. (d) Cross-trial phase consistency and slot decoding. (e,f) Time/content classifications and their stability across trial halves.}\label{fig:multiseed-robustness}
\end{figure}

Independent reporting uses the official MNIST test-image pool. Calibration uses a dedicated trial batch at $T_2=135$; neural analyses, phase decoding, and native-duration retrieval use a separate batch at the same delay. Calibration and reporting use distinct trial sampling and noise streams from the same test-image pool. Evaluation uses batch size 128 and FP32 model inference. For within-trial phase separation, we fit each present output neuron's phase over the final $T_3$ pre-cue steps using the harmonic fit in \appref{app:cells}. A trial contributes when all $N$ present-item phases are finite and each fit has $R^2\geq0.5$. Sorting these phases around the circle, including the closing gap, gives circular gaps $\Delta\phi_i$. We define
\begin{equation}
S_{\mathrm{trial}}=\frac{N}{2\pi}\min_i\Delta\phi_i.
\end{equation}
The score lies in $[0,1]$: zero indicates coincident phases and one indicates equally spaced phases. It measures separation within a trial without requiring a fixed phase-to-slot mapping across trials. Panel a plots the median valid-trial score against retrieval accuracy over all trials. Across the twelve checkpoints, test accuracy is $95.74\%$--$98.66\%$. The ten models other than a03 and a07 have valid-phase coverage of $99.41\%$--$100\%$ and median separation of $0.540$--$0.853$. For a03 and a07, separation medians are $0.256$ and $0.160$, with coverage of $9.93\%$ and $1.29\%$, respectively. Their medians therefore summarize a small subset of trials under the common $T_3$-period fit.

Panel b tests whether retrieval depends on the timing of the learned gate. We hold each trial's output-neuron activity fixed and circularly shift the normalized gating weights through every integer offset in the native $T_3$ window, then recompute the weighted readout and prediction. All trials contribute, including originally incorrect trials. Each curve represents one model. Averaging accuracy over all nonzero shifts yields $16.72\%$--$29.12\%$, compared with $95.74\%$--$98.66\%$ at zero shift. For a02, the median separation is $0.540$ and accuracy decreases from $97.49\%$ to $26.23\%$ under nonzero shifts. Visual inspection of its trial activity shows phase-separated oscillations and gate-aligned retrieval consistent with OPC.

Panel d retains the complementary cross-trial measures. Robust $R_\phi$ uses another evaluation batch at $T_2=130$ with inference retrieval extended to $4T_3$, applying the $3T_3$ window and definition in \appref{app:conditions}. Slot references are circular means of calibration-set output phases; held-out phases are assigned to their nearest reference. Macro-averaged accuracy is chance-corrected as $(\mathrm{accuracy}-1/N)/(1-1/N)$. Nine of the twelve checkpoints ($75\%$) show stable cross-trial OPC by the existing robust $R_\phi>0.9$ criterion, with slot accuracy of $99.17\%$--$99.92\%$. The a02 network provides an additional, weaker within-trial OPC example: its separation score and visually inspected output trajectories show phase-separated retrieval, although its slot accuracy is $31.47\%$ and cross-trial phase consistency is low. Thus, the nine-checkpoint count describes stable cross-trial OPC, while a02 exhibits a less consistently organized phase code.

The continued a03 and a07 checkpoints have robust $R_\phi=0.448$ and $0.180$, with slot accuracies of $66.82\%$ and $29.96\%$. Their test accuracies, $96.81\%$ and $95.74\%$, remain below the $98.54\%$ of the other five-item model, a10, whose phase code is strongly organized. These two retrained and continued models provide cases of task completion with weaker measured phase organization. Together with a10, they show variation in the quality of phase representations across the three five-item settings. The low phase-fit coverage also matters: a07 has a median output period near half its training $T_3$, whereas the common phase fit assumes period $T_3$.

Output-neuron and hidden-neuron periods use the same full-positive-frequency estimator described above, over each model's final $T_3$ pre-cue steps. Output spectra are averaged separately for each digit using only trials containing that digit; hidden spectra are averaged separately for each neuron. Across models, median output period$/T_3$ ranges from $0.486$ to $0.963$, and median hidden period$/T_3$ from $0.854$ to $0.952$. In \figref{fig:multiseed-robustness}c, the two plots share model identifiers, with training $T_3$ in parentheses; the dashed line marks a ratio of one. Output dots represent individual digit neurons; hidden distributions include every neuron. Violin widths are normalized within each model and bars mark medians. The frequency is estimated freely rather than fixed to $1/T_3$. The short pre-cue window limits frequency resolution.

\paragraph{Learned adaptation and the intrinsic rhythmic timescale.}
The period variation in \figref{fig:multiseed-robustness}c can be related to the learned adaptation gains. Across these twelve models, $\alpha_h=0.1$, $\alpha_\theta=0.01$, and $\gamma=1$ are fixed; $m_i$ is the only learned parameter among the four local dynamical parameters. Recurrent weights are also learned and determine how the units interact. To isolate the intrinsic dynamics, we hold the input and recurrent drive fixed and subtract the resulting equilibrium. Writing $a=\alpha_h$ and $b=\alpha_\theta$, the implemented update order gives
\begin{equation}
\begin{pmatrix}\delta h_{i,t+1}\\\delta\theta_{i,t+1}\end{pmatrix}
=
\underbrace{\begin{pmatrix}
1-a & -\gamma\\
bm_i(1-a) & 1-b-b\gamma m_i
\end{pmatrix}}_{A_i}
\begin{pmatrix}\delta h_{i,t}\\\delta\theta_{i,t}\end{pmatrix}.
\label{eq:intrinsic-adaptation-map}
\end{equation}
In particular, adaptation is updated using $h_{i,t+1}$. The trace is $2-a-b-b\gamma m_i$, and the determinant is $(1-a)(1-b)$. For complex-conjugate eigenvalues, their angle gives the intrinsic period in simulation steps:
\begin{equation}
P_0(m_i)=\frac{2\pi}{\displaystyle\arccos\!\left[
\frac{2-a-b-b\gamma m_i}{2\sqrt{(1-a)(1-b)}}\right]}.
\label{eq:intrinsic-adaptation-period}
\end{equation}
With the other three parameters fixed, increasing $m_i$ shortens this intrinsic period within the oscillatory regime. The eigenvalue modulus is $\sqrt{(1-a)(1-b)}\simeq0.944$, so this isolated component is damped. Sustained collective oscillations additionally involve recurrent coupling.

\begin{figure}[!t]
\centering
\includegraphics[width=\linewidth]{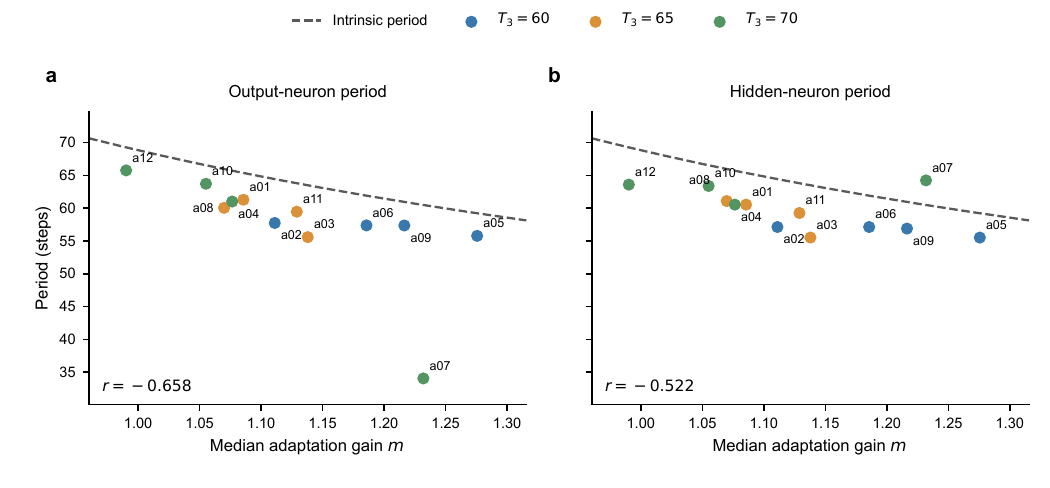}
\caption{\textbf{Learned adaptation gains track the rhythmic timescale across networks.} Each point is one of the twelve networks in \figref{fig:multiseed-robustness}c, relating its median adaptation gain to its median output-neuron (a) or hidden-neuron (b) period. Colors denote training $T_3$. The dashed curve is the intrinsic-period prediction in Eq.~\eqref{eq:intrinsic-adaptation-period}, with no fitted coefficients.}
\label{fig:adaptation-period}
\end{figure}

For each network, we extract $\widetilde m=\operatorname{median}_i m_i$ from the same checkpoint used in \figref{fig:multiseed-robustness}c, including the continued a03, a07, and a10 checkpoints. The gains span $\widetilde m=0.990$--$1.276$. Their Pearson correlations with output and hidden period medians are $-0.658$ and $-0.522$, respectively (\figref{fig:adaptation-period}). Applying Eq.~\eqref{eq:intrinsic-adaptation-period} to $\widetilde m$, without fitting additional coefficients, gives correlations of $0.659$ and $0.541$ with the measured periods. After subtracting the mean within each training-$T_3$ group, the adaptation--period correlations are $-0.826$ for outputs and $-0.175$ for hidden neurons. Mean absolute prediction errors are $6.50$ and $4.99$ steps ($13.76\%$ and $8.44\%$), respectively. The intrinsic prediction captures part of the period variation, with larger departures in the expanded cohort, particularly the faster output rhythm of a07. Collective periods also depend on learned recurrent coupling.

\begin{table}[htbp]
\centering
\caption{Training outcomes and the current analysis cohort. The a03 and a07 checkpoints follow retraining and continuation; a10 follows continuation of its original run. The remaining rows use original best checkpoints.}
\label{tab:multiseed-runs}
\small
\begin{tabular}{lrrrrrl}
\toprule
Run & $N$ & $T_3$ & $H$ & Best epoch & Val. acc. & Analysis status\\
\midrule
a01 & 4 & 65 & 550 & 53 & 0.9802 & Included\\
a02 & 4 & 60 & 550 & 54 & 0.9712 & Included\\
a03 & 5 & 65 & 500 & 253 & 0.9618 & Retrained + continued\\
a04 & 3 & 70 & 500 & 55 & 0.9827 & Included\\
a05 & 4 & 60 & 500 & 55 & 0.9799 & Included\\
a06 & 4 & 60 & 500 & 55 & 0.9803 & Included\\
a07 & 5 & 70 & 450 & 244 & 0.9519 & Retrained + continued\\
a08 & 3 & 65 & 550 & 55 & 0.9841 & Included\\
a09 & 4 & 60 & 550 & 55 & 0.9809 & Included\\
a10 & 5 & 70 & 500 & 165 & 0.9786 & Included, continued\\
a11 & 4 & 65 & 450 & 54 & 0.9798 & Included\\
a12 & 3 & 70 & 500 & 54 & 0.9835 & Included\\
\bottomrule
\end{tabular}
\end{table}

\subsection{Robustness of time and content cells}
\label{app:robust-cells}
All twelve networks use the same classification and content-priority assignment defined in \appref{app:cells}, with each model's own frozen calibration references. At the common concentration threshold $R\geq0.7$, time-cell candidates appear in all 12 models, while clearly developed content-cell populations appear in the nine models with stable cross-trial OPC. The remaining group contains neurons not assigned to either class, including poor fits. All fractions use the full hidden size $H$ as denominator, and \figref{fig:multiseed-robustness}e,f displays every model's computed counts and stability scores.

Across the nine models with both populations, exclusive time-cell counts range from 51 to 121 ($10.2\%$--$24.2\%$ of $H$), and content-cell counts from 205 to 403 ($41.0\%$--$73.3\%$). The original criteria overlap in 1--17 neurons per model. Dividing reporting trials by even versus odd index and recomputing the masks with the same calibration references gives time-cell Jaccard indices of $0.943$--$0.992$ and content-cell indices of $0.948$--$0.981$. This recurring coexistence links well-organized OPC with complementary time and content representations across independently trained networks.

The common threshold also assigns 156 time and 6 content candidates in a02, 78 and 15 in a03, and 107 and 3 in a07. These three models have only sparse content classifications at this single threshold. Their larger remaining populations may support the output readout through distributed activity patterns that are not captured by the present single-cell phase criteria. This is a possible account of their task performance, including the weaker within-trial OPC observed in a02; how the remaining neurons contribute to readout remains to be determined.

\subsection{Robustness of recurrent connectivity across seeds}
\label{app:robust-connectivity}
\begin{figure}[!t]
\centering
\includegraphics[width=\linewidth]{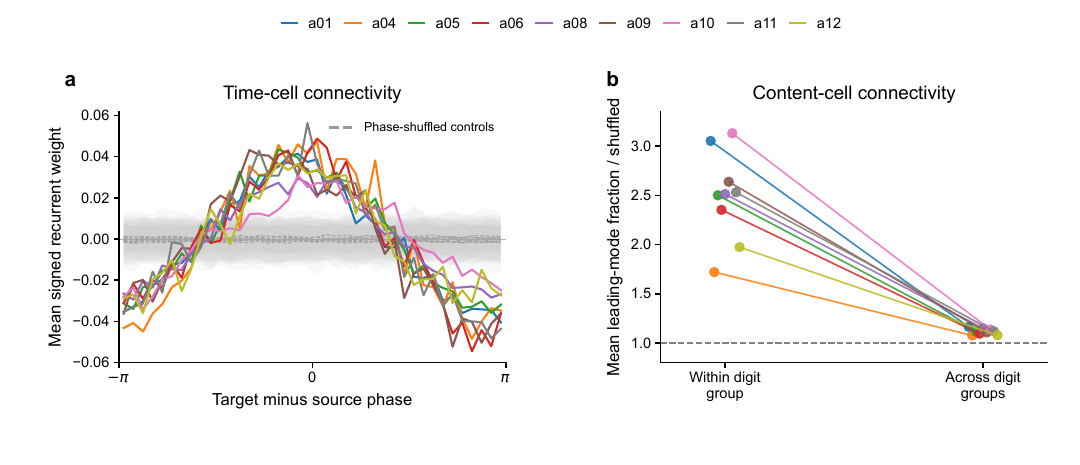}
\caption{\textbf{Time- and content-cell connectivity patterns recur across independently trained networks.} (a) Time-cell connection kernels share the phase-ring organization seen in the main model. (b) Content-cell connectivity shows consistent within-digit organization across the same nine networks. Colors identify models in both panels.}
\label{fig:multiseed-connectivity}
\end{figure}

The recurrent-connectivity patterns in \figref{fig:cells}e,f recur across the nine models with developed time and content populations (\figref{fig:multiseed-connectivity}). These are the same nine stable-OPC checkpoints identified above; their inclusion precedes the connectivity analysis. Each network contributes one connectivity kernel and one mean structure score for each content-block category.

We reuse the signed-weight analyses and shuffle controls in \appref{app:connectivity}, with each checkpoint's cell assignments and preferred phases from \appref{app:robust-cells}. Recurrent weights follow the source-to-target convention $\mathbf hW$, and self-connections are excluded. Time-cell connections are averaged in 40 equal bins of target-minus-source phase; 200 permutations of the preferred phases produce each model's shuffled mean and pointwise 5th--95th percentile band, shown in gray. The kernel is summarized by the first-harmonic fit used in \appref{app:threshold-sensitivity}. All nine kernels have a positive cosine component and show positive connections near zero phase difference and negative connections near opposite phases. The first harmonic explains $90.9\%$--$97.7\%$ of across-bin variance; correlations with the cosine are $0.886$--$0.972$.

For content cells, we apply the same block-wise centering, leading-mode fraction, and 200 within-block weight permutations as in \figref{fig:cells}f. Each model contains all ten digit groups and contributes ten within-digit and 90 directed across-digit blocks. Panel b connects each model's mean normalized structure score for the two categories. The within-digit means range from $1.720$ to $3.130$, and the across-digit means from $1.076$ to $1.164$; within-digit structure is stronger in all nine models. Across models, the corresponding means and sample standard deviations are $2.490\pm0.451$ and $1.116\pm0.031$. Together, these results support the robustness of the time- and content-cell connectivity organization in \figref{fig:cells}e,f across independently trained networks.

\section{Proposed biological test of absolute and item-relative phase tuning}
\label{app:biological-test}
The phase organization in \figref{fig:cells} predicts two neural populations in biological recordings: one tuned to absolute phase within a shared oscillatory cycle, and another tuned to phase relative to a particular memory item. Changing the item's serial position should preserve the first population's absolute phase tuning while shifting the second population's phase together with the item's reactivation. We propose the following recording experiment to test this prediction.

\paragraph{Task and recordings.}
Subjects would memorize sequences of four distinct visual items and, after a variable blank delay, report the item at a cued serial position. Repeated permutations of the same item set would place each item equally often at every position, with cue positions balanced independently. Presentation timing would be matched across permutations, and response locations would be randomized to separate item identity from motor preparation. Simultaneous single-neuron and population recordings would cover encoding, maintenance, and retrieval. The primary analysis would use the blank delay before the cue, when the target position is still unknown.

\paragraph{Phase reference and analysis.}
Calibration trials would establish whether a recurring population rhythm is present, estimate its frequency, and define a common phase reference. Here, absolute phase means phase relative to this shared rhythm. The neuron being tested would be excluded from reference construction, and the same reference rule would be applied across item identities and permutations. An item decoder trained on separate calibration trials would estimate each memory item's reactivation phase from a population subset excluding the tested neuron. Item-relative phase would be the circular difference between the neuron's phase and the decoded item's phase. Preferred items and analysis settings would be fixed on calibration trials and evaluated on held-out trials. This tests the distinction underlying the concentration measures in \appref{app:cells}, using decoded item reactivation as the biological reference. Frequency and reliability criteria would be determined from recordings rather than by converting model steps into a prescribed frequency band.

\paragraph{Predicted dissociation.}
Neurons tuned to absolute phase should retain their preferred phases across changes in item identity and order, with the population spanning the cycle. Their activity should recur over successive cycles during longer delays, distinguishing a cyclic reference from a single sequence of elapsed-time responses. Neurons with item-relative phase tuning should maintain a consistent phase offset from a particular memory item's reactivation across permutations. When that item changes serial position, the neuron's absolute phase should shift with the decoded item's phase, while their phase difference remains stable. Subtracting the item's phase should therefore increase cross-trial phase concentration for this population. The critical comparison is between invariance in the shared phase reference and invariance relative to a memory item.

\paragraph{Controls and interpretation.}
Permutation tests would shuffle item-to-position assignments within matched delay conditions, and phase effects would be evaluated alongside firing-rate changes. Cell selection and hypothesis tests would use separate trials, with uncertainty summarized across sessions and subjects. Support for the prediction would require both phase patterns to generalize to held-out permutations within the same recorded population. An absence of item-dependent phase shifts or of a stable cyclic reference would challenge this proposed organization under the tested conditions. Observing the dissociation would support the representational prediction; identifying the cellular mechanism of reciprocal coupling would require further perturbation experiments.

\end{document}